\documentclass[conference]{IEEEtran} 
\IEEEoverridecommandlockouts
\usepackage{graphicx}
\usepackage{epsfig}
\usepackage{subfigure}
\usepackage{tabularx}
\usepackage{amssymb}
\usepackage{latexsym}
\usepackage{amsmath}
\usepackage{algorithm}
\usepackage[noend]{algpseudocode}
\usepackage{color}
\usepackage{relsize}
\usepackage{amsthm}
\usepackage[,skip=3pt]{caption}
\usepackage{cite}

\newcommand{\AOIT}{TAoI }

\begin{document}

\title{Joint Age of Information and Self Risk Assessment for Safer 802.11p based V2V Networks}

%
%
%

\author{Biplav Choudhury\thanks{This research is partially supported by the Office of Naval Research (ONR) under MURI Grant N00014-19-1-2621. 

This paper has been accepted for publication in IEEE INFOCOM 2021. This is a preprint version of the accepted paper.},
        Vijay K. Shah,
        Avik Dayal, and 
        Jeffrey H. Reed \\
        Wireless@Virginia Tech \\ Bradley Department of Electrical and Computer Engineering, Virginia Tech, Blacksburg, USA \\
        Emails:\{biplavc, vijays, ad6db, reedjh\}@vt.edu
}
\maketitle

\begin{abstract}

Emerging 802.11p vehicle-to-vehicle (V2V) networks rely on periodic Basic Safety Messages (BSMs) to disseminate time-sensitive safety-critical information, such as vehicle position, speed, and heading -- that enables several safety applications and has the potential to improve on-road safety. Due to mobility, lack of global-knowledge and limited communication resources, designing an optimal BSM broadcast rate-control protocol is challenging. Recently, minimizing Age of Information (AoI) has gained momentum in designing BSM broadcast rate-control protocols. In this paper, we show that minimizing AoI solely does not always improve the safety of V2V networks. Specifically, we propose a novel metric, termed Trackability-aware Age of Information TAoI, that in addition to AoI, takes into account the self risk assessment of vehicles, quantified in terms of self tracking error (self-TE) -- which provides an indication of collision risk posed by the vehicle. Self-TE is defined as the difference between the actual location of a certain vehicle and its self-estimated location. Our extensive experiments, based on realistic SUMO traffic traces on top of ns-3 simulator, demonstrate that TAoI based rate-protocol significantly outperforms baseline AoI based rate protocol and default $10$ Hz broadcast rate in terms of safety performance, i.e., collision risk, in all considered V2V settings.

\end{abstract}

\begin{IEEEkeywords}
V2V networks, 802.11p, DSRC, Tracking Error, Collision Risk, Information Freshness, Age of Information
\end{IEEEkeywords}

%
\IEEEpeerreviewmaketitle

\section{Introduction}
\label{section - Introduction}
%
%
%
%
Vehicle-to-Vehicle (V2V) communications is envisioned as one of the key enablers for Intelligent Transportation Systems (ITS), mainly due to its potential of improving on-road safety in tomorrow's smart cities~\cite{Connecte92:online}.
The two major standards for V2V communications are the IEEE 802.11 based Dedicated Short Range Communication (DSRC) and LTE-based Cellular-V2X (C-V2X). DSRC uses 802.11p to establish a decentralized network where every vehicle transmits at fixed time intervals, usually $100$ ms~\cite{naik2019ieee}. C-V2X, on the other hand, was developed in the Release 14 of 3GPP and is a cellular technology with the additional capability of using sidelink interface PC5 for direct V2V communications. 
Newer versions of both DSRC and C-V2X are being developed, respectively called as 802.11bd and  NR-V2X~\cite{naik2019ieee}. 

The three major applications that will be supported by V2V communications are - (i) \textit{Safety Applications} - vehicles periodically broadcast their kinematics parameters (such as vehicle position, speed etc.) in messages called Basic Safety Messages (BSMs) \cite{thota2019v2v}, (ii) \textit{Cooperative Message Applications} - these are cooperative traffic efficiency messages controlled by traffic management systems with the objective of improving traffic flow \cite{bohm2011real}, and (iii) \textit{Media Sharing} - vehicles can share infotainment content in a peer-to-peer fashion \cite{9117368}. In this work, we consider a 802.11p V2V network communicating using DSRC standard where the focus is on safety applications. 
BSMs carry time-sensitive safety-critical information such as the sender vehicle's location, speed and heading, which upon reception, allows a receiving vehicle to accurately localize the nearby sender vehicles. Thus, BSMs enable several safety applications like cooperative collision warning (CCW) \cite{sengupta2007cooperative}, electronic emergency brake light (EEBL) \cite{szczurek2012estimating} and slow/stopped vehicle alert (SVA) \cite{5395777} by minimizing the number of collision risky situations. 
Therefore it is critical that the location information in a BSM is up-to-date and \textit{fresh}, as old location information doesn't provide any benefit from a safety point-of-view. To quantify this freshness of information, Age of Information ($AoI$) has been introduced as a new metric and is defined as the time elapsed since the last BSM was generated \cite{kaul2011minimizing}. In the context of safety applications and BSMs, lower $AoI$ means fresher location information which makes $AoI$ one of the key indicators to measure the performance of time-sensitive safety applications in V2V networks.

In recent years there has been a growing body of research on minimizing the $AoI$ in V2V networks (and various other communication networks) \cite{kaul2011minimizing, 6134181, ni2018vehicular, abdel2018ultra, maatouk2019minimizing, baldesi2019keep, baiocchi2017model, llatser2016vehicular, 8984230, choudhury2020experimental, vinel2015vehicle}. For instance in ~\cite{kaul2011minimizing}, the authors propose a distributed broadcast rate control algorithm for a DSRC network that minimizes system $AoI$ by iteratively minimizing the locally computed $AoI$ at each vehicle.  Refer to Sec. \ref{section - Related Work} for detailed discussion on related works. These prior works propose to minimize $AoI$ and present $AoI$ based rate control algorithms with the consideration that lower $AoI$ will always result in better safety performance in V2V networks.

However in this work, we demonstrate that minimizing $AoI$ does not always improve on-road safety performance in V2V networks. This is mainly because $AoI$ solely depends upon the time instants of the BSM generation and reception~\cite{maatouk2019age}, and does not take into account -- (i) \textit{Value of BSM information} - does the reception of a BSM at a certain receiving vehicle actually reduce the collision risk posed by the sender vehicle?, and (ii) \textit{Current knowledge of the receiver vehicle} -- what does the receiver already know about 
the sender vehicle, for example, using tracking capability (as per SAE J2945~\cite{SAE_J2945})? To counter these shortcomings of $AoI$, we propose a novel metric, termed \textit{Trackability-aware Age of Information} ($\AOIT$), that jointly considers $AoI$ and the self risk assessment of vehicles, measured as the self trackability of the vehicles. In general, \textit{trackability} refers to how accurately the location of a vehicle can be estimated by a neighbor\footnote{neighbor vehicle means vehicles within the radio transmission range.} vehicle (based on its most recently received BSM), and better trackability means better on-road safety (as shown in Sec. \ref{section - V2V_model}). However, since 802.11p V2V networks are decentralized networks, we consider that each vehicle assesses its own trackability behavior (i.e., either trackable or non-trackable) at certain pre-specified time intervals, termed \textit{measurement intervals}, and piggybacks this information in the BSM broadcasts. This allows the receiving vehicles to identify the trackability behavior of sender vehicle. Our extensive simulation experiments based on realistic SUMO traffic traces and ns-3 network simulator show that the proposed $\AOIT$ based BSM broadcast rate control protocol significantly outperforms both the baseline 10 Hz broadcast rate (up to $40\%$) and $AoI$ based rate control algorithm (up to $12\%$), in terms of improving on-road safety, i.e., collision risk reduction, in all considered V2V scenarios. Furthermore, we also witness that the packet delivery ratio is much better in case of $\AOIT$ compared to that of $AoI$, with similar or lower usage of communication channel resources. \textit{To the best of our knowledge, this is the first work that proposes to improve on-road safety of V2V networks by jointly taking into account -- the $AoI$ and self risk assessment of vehicles.}

The rest of the paper is arranged as follows: Sec. \ref{section - Related Work} discusses related works and Sec. \ref{section - V2V_model} presents the system model. Sec. \ref{section - AOI_Limitations} provides a brief background on $AoI$ followed by its limitations when considered for safety of vehicles. Sec. \ref{section - AoIT} presents an overview of proposed metric $\AOIT$, while Sec. \ref{section - Algorithm} discusses the details of $\AOIT$ aware rate control algorithm. Sec. \ref{section - Performance Evaluation} describes the results, followed by conclusions in Sec. \ref{section - Conclusion}.

\section{Related Work}
\label{section - Related Work}

Rate (or broadcast interval) control has been actively investigated as a means of reducing the $AoI$ in vehicular networks. The analytical model formulated by Baiocchi et al. in \cite{baiocchi2017model} uses the connectivity graph of the network to demonstrate the relationship of average system $AoI$ with vehicle density and broadcast intervals. Llatser et al.~\cite{llatser2016vehicular} considered a convoy of vehicles and analysed the change in $AoI$ with changing broadcast intervals and convoy size. A similar case of platooning was considered in \cite{8984230} where the European Telecommunication Standards Institute's (ETSI) Decentralized Congestion Control (DCC) was analysed in terms of $AoI$ performance - it was shown that DCC's $AoI$ can be improved by modifying its broadcast intervals to target a specific Channel Busy Ratio (CBR). In \cite{vinel2015vehicle}, Vinel et al. show that 100ms broadcast interval for Cooperative Awareness Messages (CAM), European equivalent of BSM, reduces $AoI$ better compared to mobility based CAM triggering process defined in ETSI EN 302 637-2 \cite{ETSI_EN_302}. Our work is closest to \cite{kaul2011minimizing} where a decentralized broadcast rate control algorithm was implemented that improves the system $AoI$, by iteratively minimizing the locally computed $AoI$ at each vehicle. 

In addition to rate control, other approaches that have been used to improve $AoI$ are piggybacking neighbor's information \cite{6134181}, Eigen-vector Centrality (EvC) of the topology \cite{baldesi2019keep}, optimal contention window \cite{maatouk2019minimizing}, machine learning \cite{abdel2019ultra} etc. However considering we focus on rate control, these approaches are out-of-scope with respect to our paper.

The prior works has several limitations -- (i) \textit{none of these $AoI$ based rate control protocol works evaluate how AoI improves the  on-road safety metric, i.e., collision risk reductions}. It is worth noting that very recently the authors in \cite{choudhury2020experimental} performed experimental analysis on $AoI$ and safety metrics. However, they consider a simplistic highway model which does not capture the realistic vehicle mobility behaviors. (ii) On similar lines,  \textit{most of existing work consider over-simplified vehicle mobility models}, with no lane changing or acceleration/deceleration behaviors, and finally (iii) \textit{prior work adopt a \textit{risk-neutral} approach to $AoI$} - where each vehicle is given the same importance in reducing their $AoI$. In this work, we address these limitations by proposing a novel $\AOIT$ metric and subsequently a $\AOIT$ based broadcast rate control protocol for improving on-road safety of V2V networks.

\section{System Model}
\label{section - V2V_model}

\begin{figure}[ht]
\centering
\vspace{-0.1in}
\includegraphics[scale=0.28, trim={5.5cm 1cm 0cm 5.32cm},clip, angle=0]{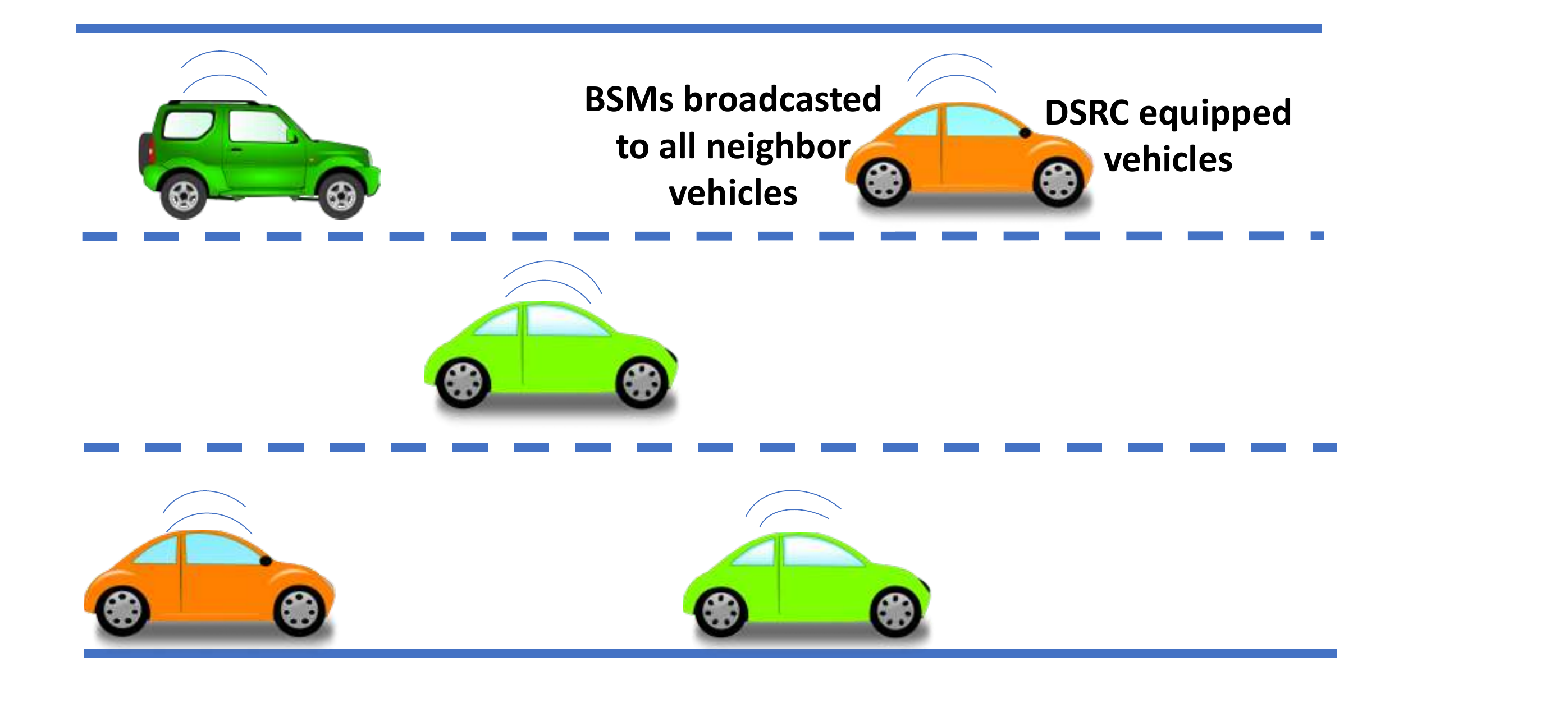}
\caption{System model: A simple 802.11p based V2V network}
\label{fig:Road Layout}
\vspace{-0.08in}
\end{figure}

Our system model consists of $\mathcal{N}$ vehicles moving in an $m$ lane rectangular road as shown in Fig. \ref{fig:Road Layout}. Each vehicle  $u \in \mathcal{N}$ moves with a certain velocity $s_u \in (0, s_{max}]$. The movement of each vehicle is implemented using the Krauss’ mobility model \cite{krauss1997metastable}, which regulates its acceleration (and deceleration); and the lane changes are integrated as per \cite{SUMOLane}. Unlike existing literature, this ensures a realistic representation of real-world driver and vehicle mobility behaviors.

Each vehicle uses $5.9$ GHz DSRC wireless technology that has been designed to support vehicular safety applications (such as CCW, EEBL and SVA) in V2V networks~\cite{Dedicate12:online}. DSRC uses existing IEEE 802.11p standard as its PHY and MAC layers; mainly because of the widespread availability of IEEE 802.11 chipsets as well as its performance and cost savings. We assume each vehicle is equipped with a single $10$ MHz DSRC radio, and the transmission power is fixed. 
Vehicular safety applications rely heavily on periodic broadcast of BSMs 
which contain the vehicle's position, heading, and other information about the transmitting vehicle. Without loss of generality, consider that each vehicle $u$ broadcasts its BSM at periodic time intervals of $\Delta$ (usually 100 ms)
which can take any random value in $[\Delta_{min}, \Delta_{max}]$. It means that there is at least $\Delta$ inter-reception delay between two consecutive BSMs from a vehicle $u$ received at a neighboring vehicle 
$v$. Moreover, since $u$ broadcasts its BSM as per Carrier Sense Multiple Access (CSMA) scheme followed by 802.11p \cite{tong2016stochastic}, it leads to additional communication delay, (i.e., queuing delay, transmission delay, and propagation delay), denoted by $x$, at receiving vehicle $v$. Because of both $\Delta$ and $x$, the information in $u$'s BSM is relatively \textit{outdated} (or stale) by the time it is received by the neighboring vehicle $v$. \textit{The lack of fresh or up-to-date information at a receiver may lead to wrong positioning of sender vehicle, i.e. $v$ will not be able to localize $u$ correctly and presents a potential on-road safety hazard}.

To address this, a promising solution is to enable a receiving vehicle $v$ to track (or \textit{estimate}) $u$'s current location during these inter BSM reception times (in time instants of no BSM receptions). As per SAE J2945 \cite{ahmad2019v2v}, a linear extrapolation based on the last known velocity, position  and heading as per the most recent BSM received can be used to estimate the sender vehicle's current location. However, note that a linear extrapolation will not always work well if the vehicle being tracked doesn't have a fixed linear mobility behavior. This will give rise to a non-negligible error at a vehicle $v$ while estimating the location of neighboring vehicle $u$, which we call \textit{Tracking error} (TE). See Sec. \ref{subsubsection - Tracking Error} for details. Lower TE value means that $v$ is able to track $u$ (or estimate $u$'s location) relatively well, and vice versa. This tracking error across all the vehicles is then used to calculate the on-road safety of V2V networks, measured as \textit{Collision risk}. See Sec. \ref{subsubsection - Collision Risk}.

\subsubsection{\underline{Tracking Error (TE), $\tau_{e, uv}$}}
\label{subsubsection - Tracking Error}
$\tau_{e, uv}$ is defined as the difference between the ground truth location of the sender vehicle $u$ and the sender's location as estimated by its neighboring vehicle $v$. Let the most recent BSM received at $v$ from $u$ was generated at time $t'$. Then at any time $t > t'$, $v$ can estimate $u$'s current location as $(\hat{x}_{uv}(t), \hat{y}_{uv}(t))$ where

\begin{equation}
    \begin{aligned}
    \hat{x}_{uv}(t) = {x}_{u}(t') + s_{u}(t')\cos(\theta_{u}(t'))\delta t \\
    \hat{y}_{uv}(t) = {y}_{u}(t') + s_{u}(t')\sin(\theta_{u}(t'))\delta t
    \end{aligned}
    \label{equation - position_prediction}
\end{equation}

\noindent where $\delta t$ = $(t - t')$, and $(x_u(t'), y_u(t')), s_u(t')$ and $\theta_u(t')$ respectively denote $u$'s location, speed, and heading information contained in the last BSM received from $u$. Then the TE $\tau_{e, uv}$ that $v$ has in tracking $u$ at time $t$ is calculated as - 

\begin{equation}
   \tau_{e, uv}(t) = \sqrt{ (x_{u}(t) - \hat{x}_{uv}(t))^2 + (y_{u}(t) - \hat{y}_{uv}(t))^2 }
    \label{equation - tracking_error}
\end{equation}

\noindent where $(x_{u}(t), y_{u}(t))$ is the ground truth location of $u$ at $t$. Note vehicle $u$ may not see the same TE in tracking vehicle $v$, i.e., $ \tau_{e, vu}(t) \neq \tau_{e, uv}(t)$, because $v$'s BSM may have been generated at different time, say $t''$ where $t'' \neq t'$, and also $v$'s location, speed and heading may be different to that of $u$. 

The average TE that $v$ has in tracking $u$ in a certain time window $T$ can be computed as 
\begin{equation}
    \tau_{e, uv}= \frac{1}{T} \int_{T} \tau_{e, uv}(t)
    \label{equation - tracking_error_formula}
\end{equation}

\textit{Significance.} The concept of TE is used to compute the Collision Risk as described later in this Sec. \ref{subsubsection - Collision Risk}.

\subsubsection{\underline{Self Tracking Error (Self-TE), $\tau_{p, u}$}}
\label{subsubsection - Self Tracking Error} 

Self-TE $\tau_{p,u}$ is the difference between the actual location of a certain vehicle $u$, and the self-estimated location of the same vehicle. It is periodically calculated after a certain time interval and this pre-specified time interval at which $\tau_{p,u}$ is (re) computed is referred to as \textit{measurement interval} ($t_{MI}$). We utilize the same linear extrapolation approach (used in TE) that uses the position, speed and heading at  $(t - t_{MI})$ to estimate the $u$'s current location as $(\bar{x}_{u}(t), \bar{y}_{u}(t))$ at time $t$ where -

\begin{equation}
    \begin{aligned}
    \bar{x}_{u}(t) = {x}_{u}(t-t_{MI}) + s_{u}(t-t_{MI})\cos(\theta_{u}(t-t_{MI}))t_{MI} \\
    \bar{y}_{u}(t) = {y}_{u}(t-t_{MI}) + s_{u}(t-t_{MI})\sin(\theta_{u}(t-t_{MI}))t_{MI} \\    
    \end{aligned}
    \label{equation - self_position_prediction}
\end{equation}

The self-TE of $u$ at time $t$, denoted by $\tau_{p,u}(t)$, is - 

\begin{equation}
    \tau_{p,u}(t) = \sqrt{ (x_{u}(t)- \bar{x}_{u}(t))^2 + (y_{u}(t)-\bar{y}_{u}(t))^2 }
    \label{equation - self_error_formula}
\end{equation}

\textit{Significance.} This concept of self TE is the basis for identifying the trackability behavior (i.e., either trackable or non-trackable) of a certain vehicle at each measurement interval $t_{MI}$. This is the first step of our proposed $\AOIT$ rate control protocol. Refer to Sec. \ref{section - AoIT} for more details.

\subsubsection{Collision Risk}
\label{subsubsection - Collision Risk}
Here we describe how we compute on-road safety of V2V networks, measured as \textit{Collision Risk} and is inspired from the risk model developed in \cite{choudhury2020experimental}. 

\begin{figure} [htb]
    \vspace{-0.20 in}
    \centering
    \includegraphics[scale=0.42, trim={6cm 6.5cm 12cm 2.3cm},clip, angle=0]{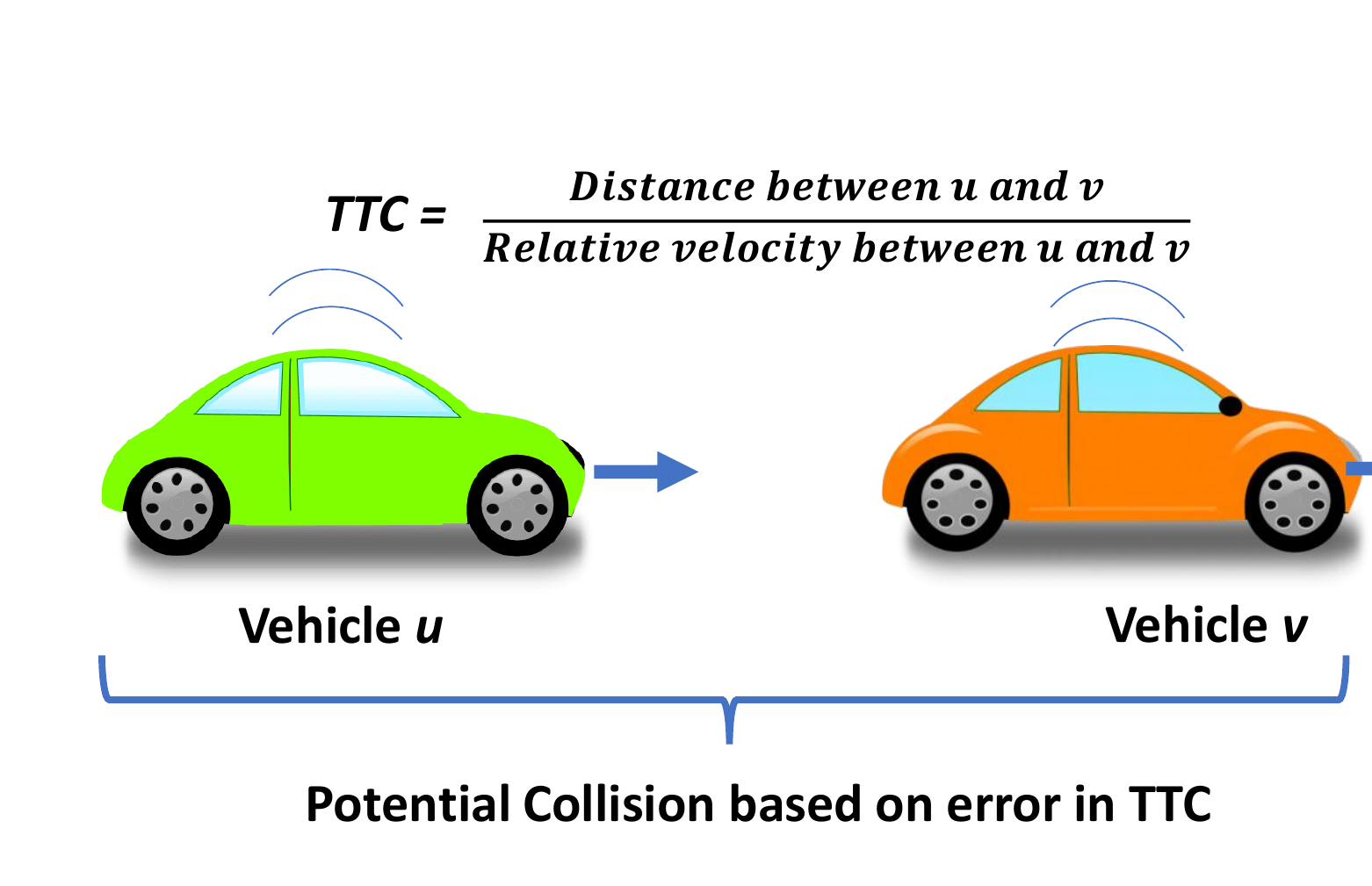}
    \vspace{-0.1in}
    \caption{TTC Calculation}
    \label{fig:Collision_Model}
    \vspace{-0.05 in}
\end{figure}

In V2V networks, Time To Collision (TTC) is an important safety metric due to which each vehicle continuously monitors the TTC to the neighboring vehicles \cite{DayalRisk, 8880565, tang2010collision, dagan2004forward}. TTC for a pair of vehicles is defined as the time needed for the distance between them to become zero, which denotes a potential collision between them. Mathematically, TTC can be computed as the ratio of distance between $u$ and $v$ and their relative velocity. 
See Fig. \ref{fig:Collision_Model}. High TTCs mean that there is no immediate threat of collision between $u$ and $v$ given the current distance and relative velocity between them and vice-versa.

Now given the TE ($\tau_{e, uv}$) that a $v$ has in tracking $u$, the error in TTC, denoted by $\delta \textit{TTC}_{uv}$, between the pair $u-v$ can be calculated as \cite{choudhury2020experimental}--

\begin{equation}
    \delta \textit{TTC}_{uv}(t) = \frac  {\tau_{e,uv}(t)}{s_{uv}(t)}
    \label{equation - TTC_error}
\end{equation}

\noindent where $s_{uv}(t)$ is the relative velocity between $u$ and $v$ at $t$. The significance of $\delta\textit{TTC}$ for collision avoidance has been studied in \cite{shladover2006analysis}. Specifically, a certain on-road situation can be considered collision risky (or dangerous) whenever the $\delta\textit{TTC}$ for any pair of vehicles exceeds the threshold value of $\delta\textit{TTC}_{thresh}$, where $\delta\textit{TTC}_{thresh}$ is the time needed for the driver to react to the possible collision and bring the vehicle to a stop. Thus, $\delta\textit{TTC}_{thresh}$ is computed as -

\begin{equation}
    \delta\textit{TTC}_{thresh} = t_{react} + t_{brake}
    \label{equation - TTC_thresh}
\end{equation}

\noindent where $t_{react}$ is 1s and refers to the time taken by the driver to respond to the situation and apply the brakes \cite{choudhury2020experimental}. $t_{brake}$ is the time taken by the vehicle to come to a complete stop after the brakes have been applied. Taking the deceleration for a vehicle as $a$ = 4.6m/s$^2$  \cite{choudhury2020experimental} makes $t_{brake} = s/a $ where $s$ is the velocity of the vehicle. A key difference from the study done in \cite{choudhury2020experimental} is that in this work, $s$ is not fixed for the entire simulation duration and changes with time for each vehicle.

Given $\delta TTC_{uv}$ and $\delta\textit{TTC}_{thresh}$, the collision risk for each vehicle pair $u$ - $v$ at any time $t$ can be computed as -

\begin{equation} \label{Collision_risk}
    CR_{uv} = 
    \begin{cases}
        1 & \delta \textit{TTC}_{uv}(t) > \delta\textit{TTC}_{thresh} \\
        0 & otherwise
    \end{cases}
\end{equation}

Using Eq. \ref{Collision_risk}, we count the number of instances between each pair of vehicles in which $\delta TTC$ exceeded $\delta\textit{TTC}_{thresh}$, as the measure of overall on-road safety of the V2V network. Note that $\delta TTC$ is directly proportional to TE, $\tau_e$, between each pair of vehicles in the network (See Eq. \ref{equation - TTC_error}). Thus, \textit{it is critical the $\tau_e$ between every pair of vehicles is minimized in order to enhance on-road safety performance of V2V networks.}

\section{Age of Information and Safety Performance}
\label{section - AOI_Limitations}
In this section, we briefly describe Age of Information ($AoI$) and then discuss the limitation of $AoI$ based rate control approach on the safety performance of V2V networks.

\subsection{Age of Information (AoI)}
\label{subsection - Age of Information}

\begin{figure} [htb]
    \centering
    \includegraphics[scale=0.25, trim={2cm 2cm 2cm 3cm},clip, angle=0]{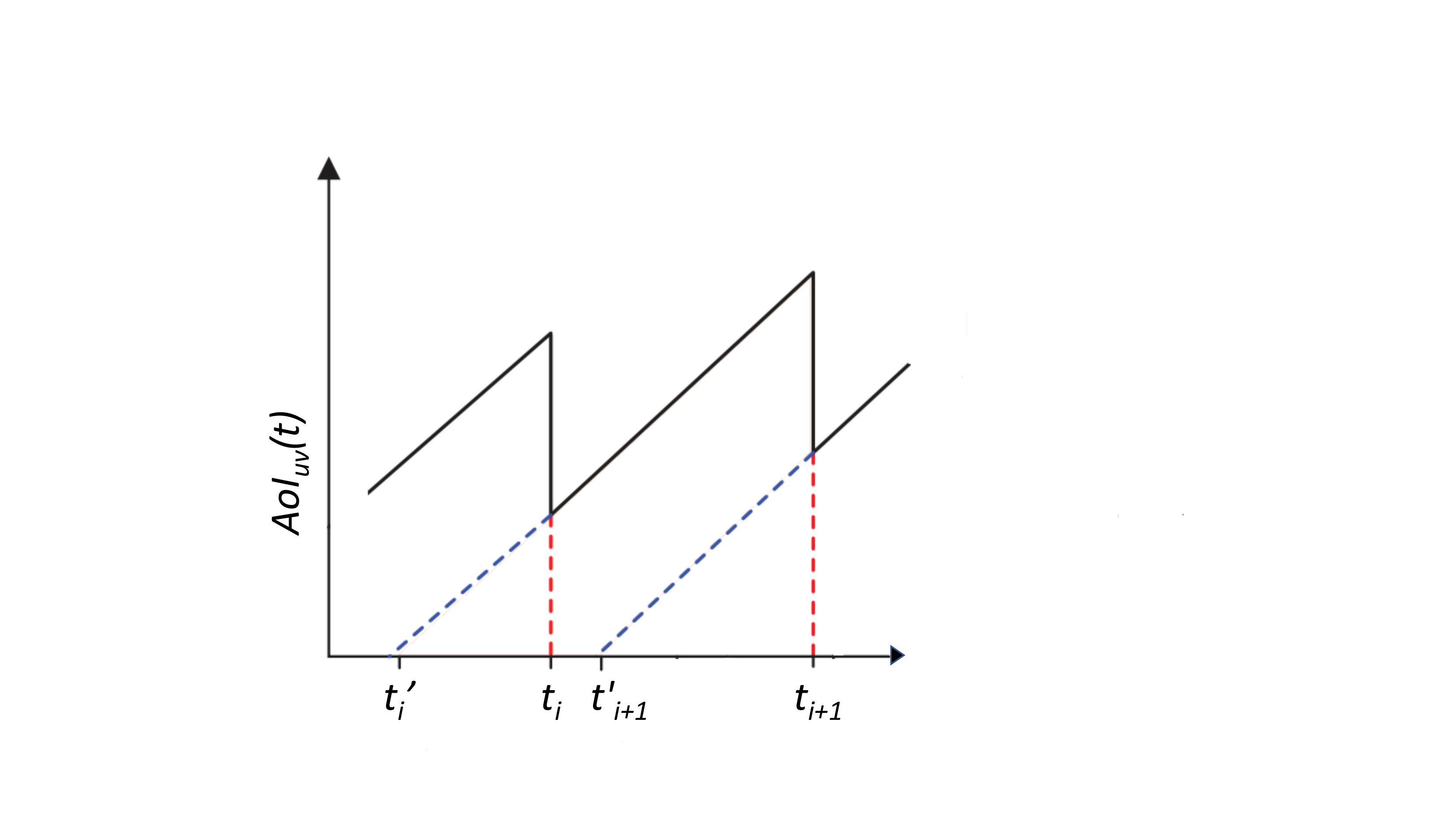}
    \caption{Evolution of $AoI_{uv}$ at $v$ based on the BSMs received from $u$. The generation and reception times of the $i^{th}$ BSM are denoted by $t'_{i}$ and $t_{i}$, which makes the overall delay for that BSM = $(t_{i} - t'_{i})$. The ${(i+1)}^{th}$ BSM is generated at $t'_{(i+1)}$, and therefore the BSM broadcast interval is given by $\Delta = (t'_{i+1} - t'_{i} )$. There is no BSM reception between $t_{i}$ and $t_{(i+1)}$, and the $AoI$ grows linearly in this interval $(t_{i+1}-t_{i})$ known as inter BSM reception interval.} 
    \label{fig:AoI plot}
    \vspace{-0.15 in}
\end{figure}

First introduced in \cite{kaul2011minimizing}, Age of Information ($AoI$) is used to 
quantify the freshness of information at a destination node (i.e. a receiving vehicle) about some measurements. 
generated by the source node (i.e. a sender vehicle). More formally, $AoI$ in V2V networks can be defined as the time elapsed since the last successfully received BSM at the receiving vehicle was generated at the sender vehicle. In order to better understand the $AoI$, let us look at Fig. \ref{fig:AoI plot} which depicts a realization of $AoI$, denoted by $AoI_{uv}(t)$, at the receiving vehicle $v$ as a function of time when a sender vehicle $u$ transmit BSMs using a first come first serve (FCFS) discipline. 
Let $t'$ denote the generation time of most recent BSM at sender vehicle $u$, then $AoI$ at receiving vehicle $v$ at current time $t$ is calculated as 

\begin{equation} \label{AoI_calculation}
    AoI_{uv}(t) = t - t'
\end{equation}

Note that $AoI_{uv}(t)$ is a zigzag-like function with a slope of $1$ between the BSM inter-reception intervals and is reset to the end-to-end delay in each time instance when $u$'s new BSM is successfully received at $v$.

Using Eq. \ref{AoI_calculation}, the average $AoI_{uv}$ between vehicle $v$ and vehicle $u$ can be calculated as the total area under the $AoI_{uv}(t)$ plot normalized by the observation interval $T$ \cite{kaul2011minimizing}, i.e. 

\begin{equation}
\label{eqn:pairwise_Avg_AoI}
    AoI_{uv} = \frac{1}{T} \int_{T} AoI_{uv}(t)
\end{equation}

Consider that $v$ has $\mathcal{N}_v \subseteq \mathcal{N}$ neighboring vehicles, i.e., $u \in \mathcal{N}_v$ where $\mathcal{N}$ is the total number of vehicles. Then the average $AoI$ at a receiving vehicle $v$ can be computed as follows.

\begin{equation}
\label{eqn:neigh_Avg_AoI}
    AoI_v = \frac{1}{|\mathcal{N}_{v}|} \sum_{u \in \mathcal{N}_{v}} AoI_{uv}
\end{equation}

Finally, the system $AoI$ with $\mathcal{N}$ vehicles can be calculated across $\mathcal{N}(\mathcal{N}-1)$ unique pairs of sender and receiver as:

\begin{equation}
\label{eqn:system_AoI}
    AoI = \frac{1}{\mathcal{N}(\mathcal{N}-1)} \sum \limits_{v \in \mathcal{N}} \sum \limits_{u \in \mathcal{N}_v} AoI_{uv}
\end{equation}

From the above discussion, it is evident that $AoI$ is less when the receiver vehicle receives BSMs frequently from the sender vehicle. As the reception of frequent BSMs means the receiver has knowledge of the most recent location of the sender, a less $AoI$ means less TE. Therefore minimizing the system $AoI$ in Eq. \ref{eqn:system_AoI} should improve the safety of all the vehicles in the network. However, when practical constraints like limited communication resources, i.e., channel capacity and different mobility behaviors of vehicles are considered, a lower system $AoI$ may not always result in improved safety performance as we shall see in Sec. \ref{subsection - AoI_Disadvatages}.

\subsection{Limitations of $AoI$ on Safety Performance}
\label{subsection - AoI_Disadvatages}

Let us consider a simple V2V scenario with two vehicles $u$ and $v$. 
For ease of understanding, we make the following assumptions for our considered V2V scenario - 

\begin{itemize}
    \item Channel capacity is limited to one BSM transmission at any time - it means at any given time $t$, only one vehicle can broadcast its BSM.
    \item There is no additional communication delay involved ( $x = 0$) - it means the inter-BSM reception intervals between a receiver $v$ and sender $u$ is equal to $\Delta$, where $\Delta$ is the BSM broadcast interval.
    \item The back-off time is taken as $\Delta$. If $u$ broadcasts at time $t = \Delta$, and another sender vehicle $v$ senses the channel busy, then $v$ will attempt its next transmission at $2 \Delta$. Thus, the instants of transmission will be $t_1$ = {$\Delta$}, $t_2$ = $2\Delta$, and so on, where either $u$ or $v$ gets to transmit. 
\end{itemize}

Under these assumptions, $AoI$ can be computed as -

\begin{equation} \label{AoI_slot}
    AoI(t + \Delta) =
    \begin{cases}
     0, & \text{if} \hspace{0.1in}  \text{a BSM is received}  \\
     AoI(t)+\Delta, & \text{otherwise}
    \end{cases}
\end{equation}

Lets consider that the mobility behavior of $u$ and $v$ is given by the following equations -

\begin{equation}
    \begin{aligned}
    y_u(t) = 2t, \quad \dot{y}_u(t) = 2 \\
    y_v(t) = t^2, \quad \dot{y}_v(t) = 2t
    \end{aligned}
    \label{equation - equation_of_motion}
\end{equation}
where $y(t)$ and $\dot{y}(t)$ are the position and velocity at time $t$ respectively. Both $u$ and $v$ will estimate each other's current locations as $\hat{y}_{vu}(t)$ and $\hat{y}_{uv}(t)$, respectively using 
Eq. \ref{equation - position_prediction}. Then the TE that $u$ has in tracking vehicle $v$ ($\tau_{e,vu}$) and $v$ has in tracking vehicle $u$ ($\tau_{e,uv}$) at time $t$ is given by Eq. \ref{equation - tracking_error}.

Let us now see how $AoI$ and an alternative rate control strategies treat $AoI$ and TE in this simple V2V scenario.

\noindent \textbf{Under $\boldsymbol{AoI}$ based rate control strategy}: both $u$ and $v$ should be allowed an equal access to the channel resources 
to broadcast their BSMs so that 
the system $AoI$ 
is minimized.
However, given the restriction of at most $1$ transmission at any time, it is intuitive that the optimal way of minimizing $AoI$ would be to allow $u$ and $v$ to broadcast their BSM at alternate intervals. E.g. if $u$ transmits at time instance $t_1$, then $v$ transmits at $t_2$ followed by $u$ at $t_3$, and $v$ at $t_4$, and so on. Lets consider $\Delta t = 1$ for the ease of presentation.

The resulting $AoI$ and TE 
is shown in Table \ref{table:c tracking AoI}. Time instants in red are $u$'s broadcast instants (and rest are $v$'s). 

From Eq. \ref{eqn:pairwise_Avg_AoI}, 
$AoI_{uv} = \frac{3}{6} = 0.5$. Similarly, $AoI_{vu} = 0.5$. Then from Eq. \ref{eqn:system_AoI}, system $AoI$ is 
$$AoI = \frac{AoI_{uv} + AoI_{vu}}{2} = 0.5$$

\noindent For TE, Eq. \ref{equation - tracking_error_formula} gives $\tau_{e,uv} = \frac{2}{6} \approx  0.33 \hspace{0.1in}, 
\hspace{0.1in} \tau_{e,vu} = \frac{15}{6}=2.5$.

\noindent \textbf{Under an alternative rate control strategy}: where $u$ transmits at $t_1$ and then $v$ gets to transmit in the remaining time. The resulting $AoI$ and TE are shown in Table \ref{table:c tracking AoIT}. The average $AoI$ and TE at $v$ are as follows -

\noindent $AoI_{uv} = \frac{15}{6} = 2.5 \text{ and } AoI_{vu} = \frac{1}{6} \approx 0.17 \Rightarrow AoI = 1.334$
Similarly, $\tau_{e,uv}= \frac{2}{6} \approx 0.33 \hspace{0.1in} \text{and} \hspace{0.1in} \tau_{e,vu} = \frac{9}{6} = 1.5$.

\noindent Regarding safety performance, the difference in TE in the two strategies occur for $\tau_{e,vu}$ at $t_4$ and $t_6$. The calculation of $\tau_{e,vu}(t_4)$ is explained here - In the $AoI$ scheme, $u$ receives a BSM from $v$ at $t_2$ when $y_v(t_2)=4, \hspace{.2cm} \dot{y}_{v}(t_2)=4$. At ${t_3}$, $\delta t = t_3-t_2 = 1$ and $u$ estimates $v$'s location as per Eq. \ref{equation - position_prediction} as $$\hat{y}_{vu}(t_3) = y_v(t_2) + \dot{y}_{v}(t_2) \times 1 = 8$$ 
TE at $u$ is calculated as $\tau_{e,vu}(t_3) = y_{v}(t_3) - \hat{y}_{vu}(t_3) = 9 - 8 = 1$. As $v$ doesn't transmit at $t_3$, $u$ again estimates $v$'s location at $t_4$ from BSM it last received at $t_2$ using Eq. \ref{equation - position_prediction} as - $$\hat{y}_{vu}(t_4) = y_v(t_2) + \dot{y}_{v}(t_2)\times2 = 12$$ where $\delta t = t_4-t_2 = 2$. Then TE $\tau_{e,vu}(t_4) = y_{v}(t_4) - \hat{y}_{vu}(t_4) = 16 - 12 = 4$.

The $\tau_{e,vu}(t_4)$ in the alternate strategy is less because $v$ gets to transmit at $t_3$ unlike in the $AoI$ strategy. Hence $u$ estimates $v$'s location at $t_4$ with $\delta t = t_4-t_3=1$ as - 
$$\hat{y}_{vu}(t_4) = y_v(t_3) + \dot{y}_{v}(t_3)\times 1 = 9+6\times1 = 15$$ and the new TE is   $\tau_{e,vu}(t_4) = y_{v}(t_4) - \hat{y}_{vu}(t_4) = 16 - 15 = 1$. In this way, $\tau_{e,vu}$ reduces from 4 to 1, thereby improving safety, even though system $AoI$ has degraded ($1.334>0.5$) when the alternative rate control strategy is used.

\begin{table}[ht]
\centering
\caption{$AoI$ rate control strategy}
\vspace{0in}
\begin{tabular}[t]{ccccccc}
\hline
Time & \textcolor{red}{$t_1$} & $t_2$ & \textcolor{red}{$t_3$} & $t_4$ & \textcolor{red}{$t_5$} & $t_6$ \\
\hline
$AoI_{uv}(t)$   & 0 & 1 & 0 & 1 &  0 &  1 \\ \label{AoI_case1}
$y_u(t)$        & 2 & 4 & 6 & 8 & 10 & 12 \\
$\hat{y}_{uv}(t)$  & 0 & 4 & 6 & 8 & 10 & 12 \\
$\tau_{e,uv}(t)$& 2 & 0 & 0 & 0 &  0 & 0 \\ \hline
$AoI_{vu}(t)$   & 1 & 0 & 1 & 0 &  1 &  0 \\
$y_v(t)$        & 1 & 4 & 9 & 16 & 25 & 36 \\
$\hat{y}_{vu}(t)$  & 0 & 0 & 8 & 12 & 24 & 32 \\
$\tau_{e,vu}(t)$& 1 & 4 & 1 & \textbf{4} &  1 & \textbf{4} \\ \hline
\label{table:c tracking AoI}
\end{tabular}
\vspace{-0.2in}
\end{table}%

\begin{table}[ht]
\centering
\caption{Alternative rate control strategy}
\vspace{0in}
\begin{tabular}[t]{ccccccc}
\hline
Time & \textcolor{red}{$t_1$} & $t_2$ & $t_3$ & $t_4$ & $t_5$ & $t_6$ \\
\hline
$AoI_{uv}(t)$   & 0 & 1 & 2 & 3 &  4 &  5 \\
$y_u(t)$        & 2 & 4 & 6 & 8 & 10 & 12 \\
$\hat{y}_{uv}(t)$  & 0 & 4 & 6 & 8 & 10 & 12 \\
$\tau_{e,uv}(t)$& 2 & 0 & 0 & 0 &  0 & 0 \\
\hline
$AoI_{vu}(t)$   & 1 & 0 & 0 & 0  &  0 &  0 \\
$y_v(t)$        & 1 & 4 & 9 & 16 & 25 & 36 \\
$\hat{y}_{vu}(t)$  & 0 & 0 & 8 & 15 & 24 & 35 \\
$\tau_{e,vu}(t)$& 1 & 4 & 1 &  \textbf{1} &  1 &  \textbf{1} \\ \hline
\label{table:c tracking AoIT}
\end{tabular}
\vspace{-0.2in}
\end{table}%

Based on these discussions, we can make the following key observations -- even though $u$'s $AoI$ at $v$ ($AoI_{uv}$) is lower in $AoI$ based rate control strategy compared to that of the alternative one, $v$ incurs the same TE ($\tau_{e,uv}$) in tracking $u$ in both the strategies. It means minimizing $AoI_{uv}$ in the $AoI$ based rate control strategy did not help in improving $u$'s safety. On the other hand, $AoI_{vu}$ is lower in alternative rate control strategy and it leads to a significant reduction in TE, $\tau_{e,vu}$, which will promisingly improve safety of $v$.

This is because, as shown in Eq. \ref{equation - equation_of_motion},  $u$ does not change its mobility behavior (i.e., speed and heading), or in other words, \textit{$u$ is trackable}. It means, after $v$ receives the first BSM from $u$ at $t_1$, $v$ accurately estimates $u$'s current location at times $t_2,\dots t_6 $ and thus, leading to $\tau_{e, uv} = 0$ for the remaining time, even in the absence of fresh BSM receptions. This fact is being harnessed by the alternative rate control strategy where after allotting a single transmission for $v$, all the remaining times ($t_2, \dots t_6$) are allotted to $v$. As $v$'s mobility behavior keeps changing with time (i.e., \textit{$v$ is non-trackable}), location estimation
doesn't perform well in tracking $v$ due to which frequent BSM receptions are required at $u$ from $v$ to enable better tracking of $v$ so that $\tau_{e,vu}$ is reduced.

Because the non-trackable vehicles have high TE and as a result are likely to have a higher value of collision risk (Sec. \ref{subsubsection - Collision Risk}), we call these \textit{risky} vehicles. Similarly, the trackable vehicles are called \textit{non-risky} vehicles.

From the above V2V scenario (and key observation), it is clear that minimizing $AoI$ does not always improve the safety of V2V networks. It becomes critical to take into account the trackability of vehicles. With that, the following deductions can be made about the relation between $AoI$ and safety in the context of vehicular networks - 

\begin{itemize}
    \item $AoI$ minimization is not important for non-risky vehicles that can be tracked well by its neighbors. 
    \item For risky vehicles, $AoI$ minimization will improve the TE, and thus, on-road safety of the network.
\end{itemize}


\section{Trackability aware AoI (\AOIT) metric}
\label{section - AoIT}

To address the shortcomings of $AoI$, we propose a novel \textit{Trackability aware AoI} ($\AOIT$) metric that accounts for both $AoI$ and trackability (i.e., risky or non-risky) of neighboring vehicles for improved safety performance of V2V networks. To get the trackability information about the neighboring vehicles, a certain vehicle $v$ can compute the TE $\tau_{e,uv}$ for each neighbor $u$ at instants of BSM receptions using Eq. \ref{equation - position_prediction} and Eq. \ref{equation - tracking_error}. However, this TE approach has two major limitations --  
(i) \textit{a vehicle learns the value of TE only at BSM reception instants}, where the inter-BSM reception time intervals may be very large (100's of milliseconds), and (ii) \textit{additional communication overheads in V2V networks}, because the receiver vehicle which computes TE, has to send back the feedback (with TE information) to every neighboring vehicle. 

A better approach to measure trackability (or riskiness) is to do \textit{self risk assessment} by utilizing self TE $\tau_{p,v}$ as vehicles having high self TE are likely to have high TE. It is computed locally at each vehicle, and does not suffer from the aforementioned limitations. Each vehicle identifies itself as a \textit{non-risky} or \textit{risky} vehicle based on its self-TE value, and shares this information with neighboring vehicles by piggybacking it along with its BSM. Based on this information, every vehicle is aware of which of the neighborhood vehicles are risky. 

For a sender-receiver pair $u$-$v$, calculation of the instantaneous and average $\AOIT$ is similar to the calculation of $AoI$. At any time $t$, $\AOIT_{uv}$ at $v$ with respect to $u$ is given by 

\begin{equation}
    \label{eqn:AoIT_defn}
    \AOIT_{uv}(t) = AoI_{uv}(t)\times \mathbf{I}(\tau_{p,v} \ge \tau_{p,th})
    \end{equation}

\noindent where $AoI_{uv}(t)$ is $AoI$ at $v$ with respect to $u$ and is computed using Eq. \ref{AoI_calculation}. $\mathbf{I}(\tau_{p,v} \ge \tau_{p,th})$ is an indicator function that equals 1 only when the self TE $\tau_{p,v}$ exceeds a fixed threshold $\tau_{p,th}$ and is 0 otherwise. Therefore, $\AOIT$ takes on the value of $AoI$ when $\tau_{p,v}$ is greater than $\tau_{p,th}$, otherwise it is 0. 

Similarly, the average $\AOIT$ for the pair $u$-$v$ over the observation interval $T$ is calculated as:

\begin{equation}
\label{eqn:pairwise_Avg_AoIT}
    \AOIT_{uv} = \frac{1}{T} \int_{T}\AOIT_{uv}(t)
\end{equation}

\noindent Then the average $\AOIT$ at $v$ becomes -


\begin{equation} \label{AoIT_average}
    \AOIT_{v} = \frac{1}{|\mathcal{N}^{r}_{v}|}\sum_{u \in \mathcal{N}^{r}_{v}} \AOIT_{uv}
\end{equation}

\noindent where $\mathcal{N}^r_v \subseteq \mathcal{N}_v$ is the set of risky neighboring vehicles of $v$ and can be computed as follows -
\begin{equation} \label{risky_neighbor_list}
\mathcal{N}_{v}^r = \{u|\textbf{I}(\tau_{p,u} \ge \tau_{p,th}), u \in \mathcal{N}_{v}\}
\end{equation}

\noindent Finally the system $\AOIT$ across all the vehicle pairs is - 

\begin{equation}
\label{eqn:system_AoIT}
    \AOIT = \frac{1}{\mathcal{N}(\mathcal{N}-1)} \sum \limits_{v \in \mathcal{\mathcal{N}}} \sum \limits_{u \in \mathcal{N}_{v}^r} \AOIT_{uv}
\end{equation}

In the example presented in Sec. \ref{subsection - AoI_Disadvatages}, we briefly discuss how $\AOIT$ based rate control strategy 
will always reduce the TE of each neighboring vehicle (whether risky or not), and thus, is a promising metric for improved on-road safety.
\begin{itemize}

    \item The value of $\AOIT_{vu}(t)$ at vehicle $u$ is $0$ because $u$ identifies itself a non-risky vehicle, i.e., $\mathbf{I}(\tau_{p,u} \ge \tau_{p,th}) = 0$ for $\tau_{p,th} > 0$. Therefore it does not need to transmit BSMs frequently (or at lower broadcast interval $\Delta_u$) to reduce its $\AOIT_{vu}$. This doesn't harm $u$'s safety because once $v$ receives $u$'s BSM, then $v$ learns $u$'s non-riskiness behavior; and the obtained $u$'s information (i.e., location, speed, and heading) contained in BSM will allow $v$ to track it accurately. 
    Here, $u$ only needs to broadcast fresh BSM only once unless it changes its mobility behavior.

    \item At vehicle $v$, $TAoI_{uv}(t) = AoI_{uv}(t)$ because $v$ identifies itself as a risky vehicle, i.e., $\mathbf{I}(\tau_{p,v} \ge \tau_{p,th}) = 1$. Here, $v$ must broadcast its BSM as frequently as possible to minimize its $TAoI_{uv}$ (or $AoI_{uv}$). This will reduce $\tau_{e, uv}$ and ensures $v$'s safety. Note that since $u$ does not need to transmit unless its mobility behavior is changed, $v$ can utilize the additional channel resources to transmit more frequently and thus, further reducing $TAoI_{uv}$ (or $AoI_{uv}$), and improving the overall on-road safety of the network. 
\end{itemize}

To summarize, $\AOIT$ metric harnesses the capability of tracking or estimation (using linear extrapolation as per SAE J2945) and enables a non-risky vehicle to broadcast its BSM with much lower priority, i.e, at higher inter broadcast intervals $\Delta$, and a risky vehicle with higher priority, i.e., at much lower $\Delta$. This will greatly reduce the TE corresponding to risky vehicles without compromising the TE of the non-risky vehicles, and thus improve the overall safety of V2V networks.

\noindent \textbf{Problem Overview.}  The problem of designing an optimal rate control strategy that minimizes system $\AOIT$ can be formulated as an Integer Linear Programming (ILP) problem.

At each time $t \in T$, let $C(t)$ denote the total channel capacity in V2V networks and $\mathcal{N}^r_v(t)$ denote the list of risky vehicles in the neighborhood of any vehicle $v$. 
Note, $\mathcal{N}^r_v(t)$ does not include $v$'s non-risky neighbors, for which $\mathbf{I}(\tau_{p,u} \ge \tau_{p,th}) = 0$ where $u \in \mathcal{N}_v$. Finally, let $r_v(t)$ denote the rate at which $v$'s broadcasts BSMs, and it can be calculated as $r_v(t) = \frac{1}{\Delta_v(t)}$. Here $\Delta_v(t)$ is the inter broadcast interval of $v$ at time $t$. Since $\Delta \in [\Delta_{min}, \Delta_{max}]$, $r_v \in [r_{min}, r_{max}]$.

\textit{Objective function.} The objective is to minimize the overall system $\AOIT$ across all vehicles over the entire time period.

\begin{align}
    &\!\min_{r_{v}(t)} \sum_{t \in T} \frac{1}{\mathcal{N}(\mathcal{N}-1)}  \sum_{v \in \mathcal{N}} \sum_{u \in \mathcal{N}_{v}^r}  TAoI_{uv}(t) \label{eq:optProb}\\
    &\text{subject to} \sum_{{u \in (\mathcal{N}_{v}}\cup v)}{r_{u}(t) < C(t)},  \forall v \in \mathcal{N}, t \in T \label{eq:constraint1}\\
    &  r_{min} \leq r_{v}(t) \leq r_{max}, \forall v \in \mathcal{N}, t \in T \label{eq:constraint2}
\end{align}

\textit{Constraints.} Eq. \ref{eq:constraint1} constrains that the sum total of broadcast rate of all vehicles in the vicinity must be less than the total channel capacity. Eq. \ref{eq:constraint2} restricts the broadcast rate between the minimum and maximum allowable rates.

Intuitively, the above ILP formulation will provide a centralized optimal solution to the $\AOIT$ minimization problem. However, it is highly impractical in our considered 802.11p based V2V networks because of \textit{lack of global information.} The exact calculation of system $\AOIT$ at any given time $t$, given in Eq. \ref{eq:optProb}, requires knowledge of TE $\tau_{e,uv}$ for all vehicle pairs in the network. However, each vehicle $v$ can only estimate TE $\tau_{e, uv}$ for its neighboring vehicles based on the BSMs it receives from those vehicles. Furthermore, it assumes the \textit{knowledge of vehicle's mobility and channel information at future time instants}, which is largely impossible to obtain in any time-varying network, including V2V networks. 

Thus, we propose a novel $\AOIT$ based rate control algorithm that attempts to minimize system $\AOIT$ of V2V networks in a decentralized manner, as shown in Sec. \ref{section - Algorithm}.

\section{$\AOIT$-based Rate Control Algorithm}
\label{section - Algorithm}

This section presents the decentralized $\AOIT$ rate control algorithm, which allows each vehicle to determine its broadcast time interval $\Delta$ (where $\Delta \in [\Delta_{min}, \Delta_{max}]$) such that the locally measured average $\AOIT$ is minimized. Our algorithm is run at each vehicle at every measurement interval $t_{MI}$. However, note that $t_{MI}$ does not need to be synchronized across all the vehicles in the network. 

The algorithm operates in two steps -- first, each vehicle identifies itself as a risky or non-risky vehicle by computing its self TE, and shares this information with the neighboring vehicles. Second, based on this knowledge of the risky and non-risky vehicles in their neighborhood, each vehicle adapts its broadcast interval $\Delta$ so as to minimize the locally measured average $\AOIT$. The details of the algorithm are as follows:

\noindent \textbf{\textit{Step 1: Vehicle Self-Risk Assessment.}} As depicted in Algorithm \ref{algorithm - 1}, each vehicle $v$ utilizes Eq. \ref{equation - self_error_formula} and calculates its self-TE $\tau_{p,v}$, based on its actual and self-estimated location every $t_{MI}$ (See line \ref{algo1_STE}). Depending upon on the self-TE value, $v$ identifies itself as a non-risky vehicle (i.e., $\mathbf{I}(\tau_{p,v} \ge \tau_{p,th}) = 0$) or a risky vehicle (i.e., $\mathbf{I}(\tau_{p,v} \ge \tau_{p,th} = 1$), and sets a \textit{self riskiness} flag respectively as $0$ or $1$. See lines (\ref{algo1_set_flag_start} - \ref{algo1_set_flag_end}). Following this, $v$ piggybacks this information in all its BSMs broadcasts in line \ref{algo1_broadcast} so that the neighboring vehicles are aware of $v$'s self riskiness behavior. As the vehicle's riskiness (risky or non-risky) behavior may change over time, as also previously discussed, the entire algorithm (including this step) is (re) run at the beginning of every $t_{MI}$.

\begin{algorithm}  
\small
	\textbf{Input:} At vehicle $v$ under consideration at time $t$ - actual location $(x_{v}(t), y_{v}(t))$, self-estimated location $(\bar{x}_{v}(t), \bar{y}_{v}(t))$
	
	\textbf{Output:} Set the flag for riskiness as 1 or 0.
	
	\begin{algorithmic} [1]
    \State Calculate the Self Tracking Error $\tau_{p,v}$ using $(x_{v}(t), y_{v}(t))$ and $(\bar{x}_{v}(t), \bar{y}_{v}(t))$ based on Eq .\ref{equation - self_error_formula}.
    \label{algo1_STE}

    \If {$\mathbf{I}(\tau_{p,v} \ge \tau_{p,th}) == 1$}
    \label{algo1_set_flag_start}
        \State Mark $v$ as a risky vehicle
       \State Set \textit{riskiness flag} = $1$ 
    \Else
        \State Mark $v$ as a non-risky vehicle
        \State Set \textit{riskiness flag} = $0$ \label{algo1_set_flag_end}
    \EndIf
    \State Piggyback the \textit{riskiness flag} in all the BSMs broadcast between current $t_{MI}$ and next $t_{MI}$.
    \label{algo1_broadcast}
    \label{algo_1_reset}
	\end{algorithmic}  
	\caption{Self Assessment of riskiness at node $v$}
	\label{algorithm - 1}
\end{algorithm}

\vspace{-0.2in}
\begin{algorithm} 
\small
	\textbf{Input:} 
	Average of the broadcast intervals of all neighbor vehicles ${\Delta_{avg}}$, $\AOIT_{v}$ of the previous $t_{MI}$ denoted by $\AOIT_{v}'$ , action taken in the previous $t_{MI}$ $\Omega$, broadcast interval change factor $\beta$, previous broadcast interval ${\Delta'_{v}}$.\\
	\textbf{Output:} New broadcast interval ${\Delta_v}$ for node $v$
    \begin{algorithmic} [1]
    \State Calculate $AoI_v$ and $\AOIT_{v}$ for the current $t_{MI}$ using Eq. \ref{eqn:neigh_Avg_AoI} and Eq. \ref{AoIT_average} respectively.
    \label{algo2_TAOI}

    \If {$AoI_v > 2\Delta_{avg}$} \Comment{if channel is congested} \label{algo_2:3_start}
        \State  {$\boldsymbol{\alpha_v}$ = INCR } \label{algo_2:3_end} 
    \EndIf

    \If{$\textbf{I}(\tau_{p,v} \ge \tau_{th}) == 0$} \Comment{If $v$ is non-risky}
    \label{algo2_non_risky_start}
    \State  {$\boldsymbol{\alpha_v}$ = SAME } \label{algo2_non_risky_end}
    
    \ElsIf{$\textbf{I}(\tau_{p,v} \ge \tau_{th}) == 1$} \Comment{If $v$ is risky}
        \label{algo2_risky_start}
    
    \If {$|\mathcal{N}^{r}_{v}| == 0$} \Comment{no risky neighbors}\label{algo_2:4_start} 
            \State {$\boldsymbol{\alpha_v}$ = DECR} 
        \label{algo_2:4_end}
            
    \ElsIf {$\AOIT_{v} < \AOIT_{v}'$} \Comment{if $\AOIT_{v}$ has improved}\label{algo_2:5_start}
        \State {new action $\boldsymbol{\alpha_v}$ = $\boldsymbol{\Omega}$} 

    \ElsIf {$\AOIT_{v} > \AOIT_{v}'$}  \Comment{if $\AOIT_{v}$ has degraded}
        \State {new action $\boldsymbol{\alpha_v}$ = $\boldsymbol{\Omega}^c$} \label{algo_2:5_end}

    \ElsIf{$\AOIT_{v} == \AOIT_{v}'$} \label{algo_2:6_start} \Comment{no change in $\AOIT_{v}$}
        \State {new action $\boldsymbol{\alpha_v}$ = SAME} \label{algo_2:6_end} 
    \EndIf
    \EndIf
    
    \If {$\boldsymbol{\alpha_v}$ == INCR} \label{algo_2:7_start}
        \State{$\Delta_v = \beta \Delta'_{v}$}
    \ElsIf {$\boldsymbol{\alpha_v}$ == DECR}
        \State{$\Delta_v = \frac{\Delta'_{v}}{\beta}$}
    \ElsIf {$\boldsymbol{\alpha_v}$ == SAME}
        \State{$\Delta_v = \Delta'_{v}$} \label{algo_2:7_end}
        
    \EndIf

    \State return $\Delta_v$ 
	\end{algorithmic}  
	\caption{$\AOIT$-based Rate Control at vehicle $v$}
	\label{algorithm - 2}
\end{algorithm}

\noindent \textbf{\textit{Step 2: Broadcast Rate Adaptation.}} 
In this second step, our algorithm iteratively minimizes the locally measured average $\AOIT_{v}$ at each vehicle $v$ by adapting the broadcast interval, i.e., $\Delta_v$ 
at each $t_{MI}$ through one of the following three possible actions\footnote{Note that the concept of utilizing these actions for rate adaption in inspired by the $AoI$ rate control algorithm proposed in \cite{kaul2011minimizing}.} -- (i) \textbf{DECR} - decrease broadcast interval, (ii) \textbf{INCR} - increase broadcast interval, and (iii) \textbf{SAME} - maintain broadcast interval. The pseudocode for this step is presented in Algorithm \ref{algorithm - 2}.

As shown in lines \ref{algo_2:3_start}-\ref{algo_2:3_end}, each vehicle $v$ first checks for channel congestion, and the congestion criteria is adopted from Kaul et al. \cite{kaul2011minimizing}. If the channel is congested, $v$ will unequivocally increase its broadcast interval $\Delta_v$ to avoid congesting it further. Then lines \ref{algo2_non_risky_start}-\ref{algo2_non_risky_end} check if $v$ is non-risky, for which $\Delta_v$ is unchanged. Otherwise $v$ is a risky vehicle (line \ref{algo2_risky_start}), and the special case where $v$ has no risky neighbors is handled in lines \ref{algo_2:4_start}-\ref{algo_2:4_end} : $v$ will reduce its $\Delta_v$ as a risky vehicle should transmit its BSMs frequently. The lines \ref{algo_2:5_start} - \ref{algo_2:5_end} forms the core of the algorithm where $\AOIT_v$ for the current and previous $t_{MI}$ are compared. If $\AOIT_v$ has improved based on the previous action, the previous action is repeated. 
Otherwise, the complimentary of the previous action is selected, where INCR and DECR are compliments of each other. If there is no change in $\AOIT_v$, action chosen is SAME as shown in lines \ref{algo_2:6_start} - \ref{algo_2:6_end}. Finally, lines \ref{algo_2:7_start}-\ref{algo_2:7_end} are used to calculate the new broadcast interval based on the action selected in the above steps. The factor by which the intervals change if the action was INCR or DECR is given by $\beta$. 
The new broadcast interval returned is maintained until the end of current $t_{MI}$.

Note there are three controlling parameters in the $\AOIT$ based rate control algorithm, as discussed below.

\begin{itemize}
    \item \textit{Interval change factor ($\beta$).} $\beta$ is used to calculate the new broadcast interval from the previous broadcast interval. When intervals are increased (action is \textbf{INCR}), the new interval is obtained by multiplying the old interval with $\beta$ and vice-versa. We set $\beta$=1.1 as per     \cite{kaul2011minimizing}.
    
    \item \textit{Measurement Interval ($t_{MI}$).} $t_{MI}$ is set at 1s. It can be higher or lower depending on how frequently the network changes. Even though $t_{MI}$=1s is a low value for evaluating $\AOIT$ in a network as a vehicle may not have received enough BSMs to get a correct estimate of its local $\AOIT$, smaller values of $t_{MI}$ helps the network to minimize $AoI$ (and hence $\AOIT$) faster \cite{kaul2011minimizing}. 
    
    \item \textit{Self TE threshold $\tau_{p,th}$.} As per the recommendations made in the SAE-J2945 \cite{SAE_J2945}, we set $\tau_{p,th}$ = 0.5m. However for practical considerations, this can be adjusted to reflect how accurate the location information needs to be, or in other words, what is the threshold for being classified as a risky vehicle. For $\tau_{p,th}$ = 0, $\AOIT=AoI$.
\end{itemize}
The calculation of the optimal $\beta$ and $t_{MI}$ are beyond the scope of this study and will investigated as a part of our future work.

\textbf{Time Complexity.} Intuitively enough, Algorithm \ref{algorithm - 1} has a running time complexity of $O(1)$. The time complexity of the Algorithm \ref{algorithm - 2}
is $O(\mathcal{N})$.
This is because the algorithm computes $\AOIT$ at vehicle $v$ for each of the neighboring vehicles, which may be the total number of vehicles in the worst case scenario. Thus, the total time complexity of $\AOIT$ based rate control algorithm is $O(\mathcal{N} + 1) = O(\mathcal{N})$.

Note the V2V network is a time-varying network where the neighborhood observed by each vehicle keeps changing over time. Because of this, the algorithm will keep changing its $\Delta_v$ for each vehicle $v$ based on the new value of local $\AOIT_{v}$. This means that there may not be a broadcast interval that is safety-optimal at all times. Hence we do not pursue convergence analysis.

\section{Performance Evaluation}
\label{section - Performance Evaluation}

This section evaluates the $\AOIT$ based rate protocol against the baseline $AoI$ based rate protocol~\cite{kaul2011minimizing} and standard $10$ Hz.

\begin{figure*}
\vspace{-0.05in}
\subfigure[\label{fig:Road_SUMO}]{
\epsfig{figure=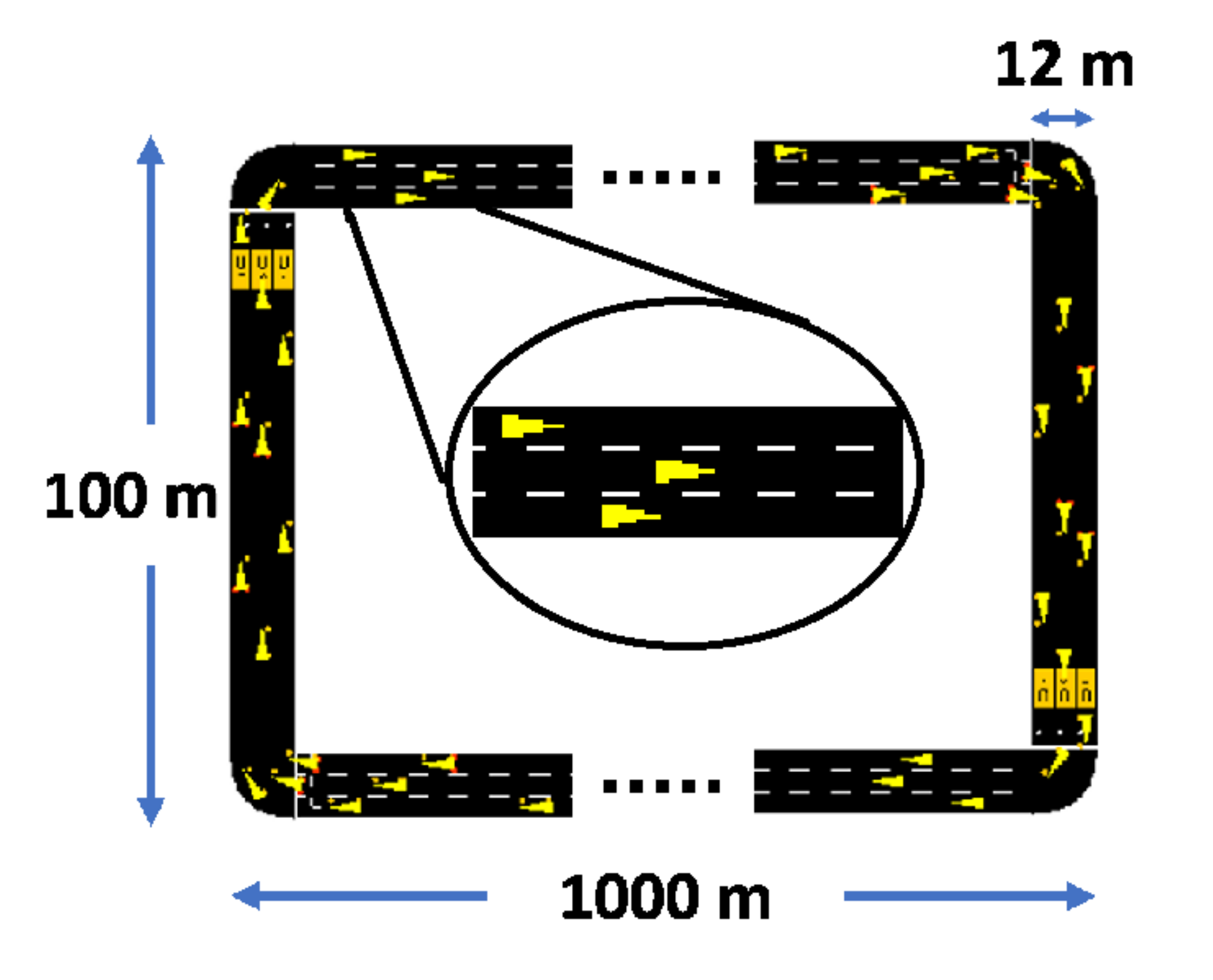, width=1.7 in}}
\subfigure[\label{fig:150U_Risk}]{
\epsfig{figure=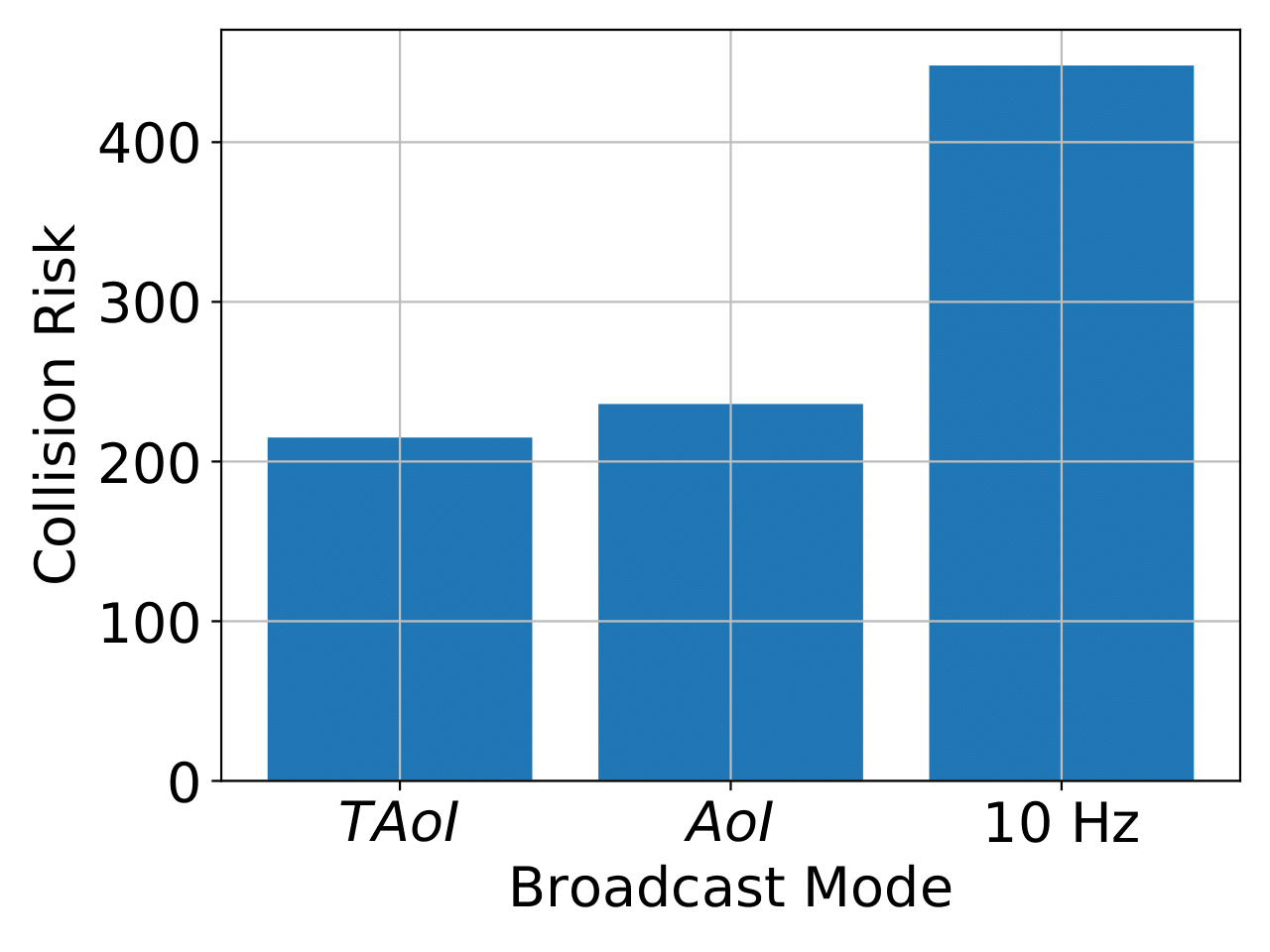,width=1.65 in}}
\subfigure[\label{fig:150U_Intervals}]{
\epsfig{figure=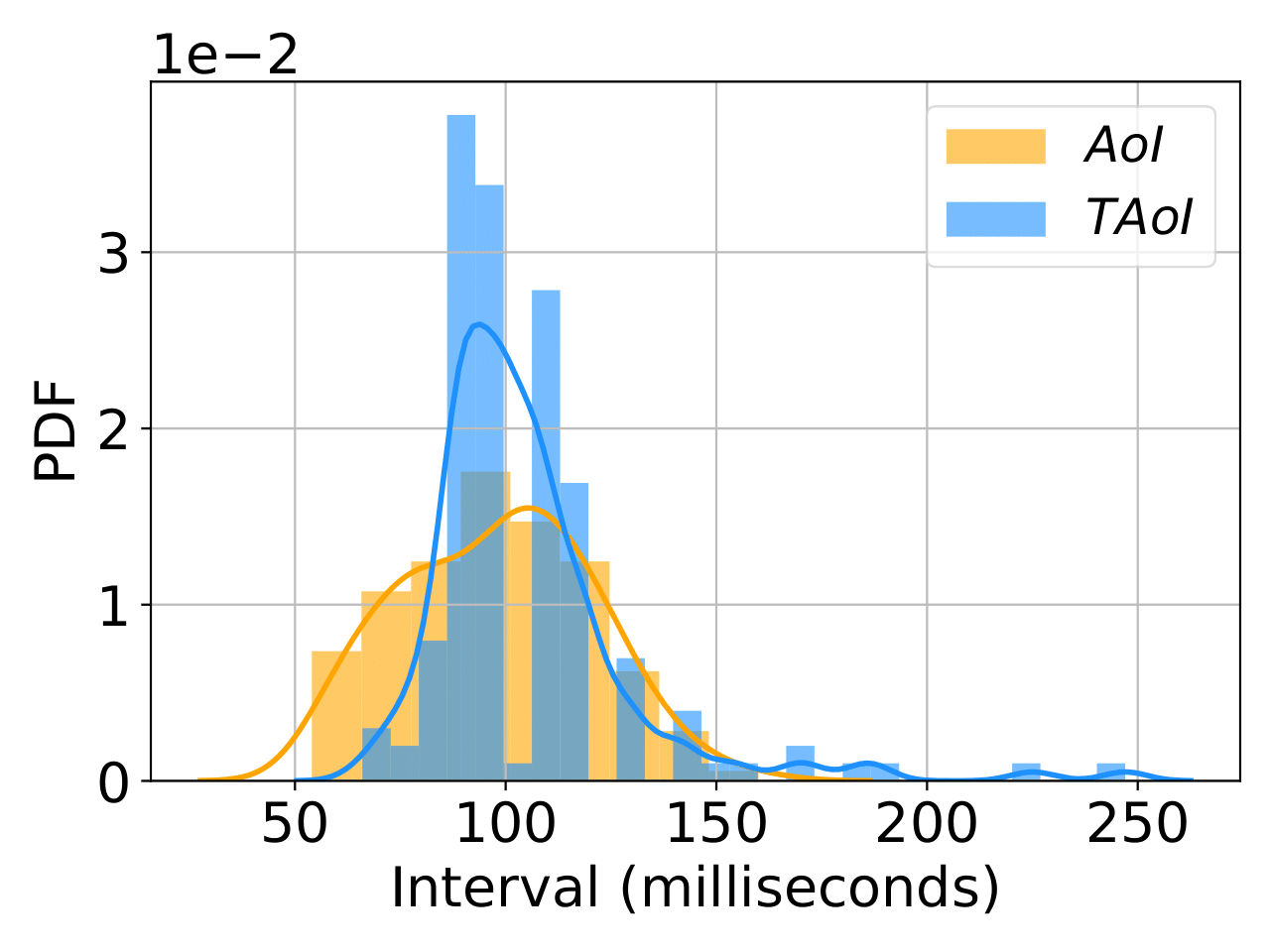,width=1.65 in}}
\subfigure[\label{fig:150U_PDR}]{
\epsfig{figure=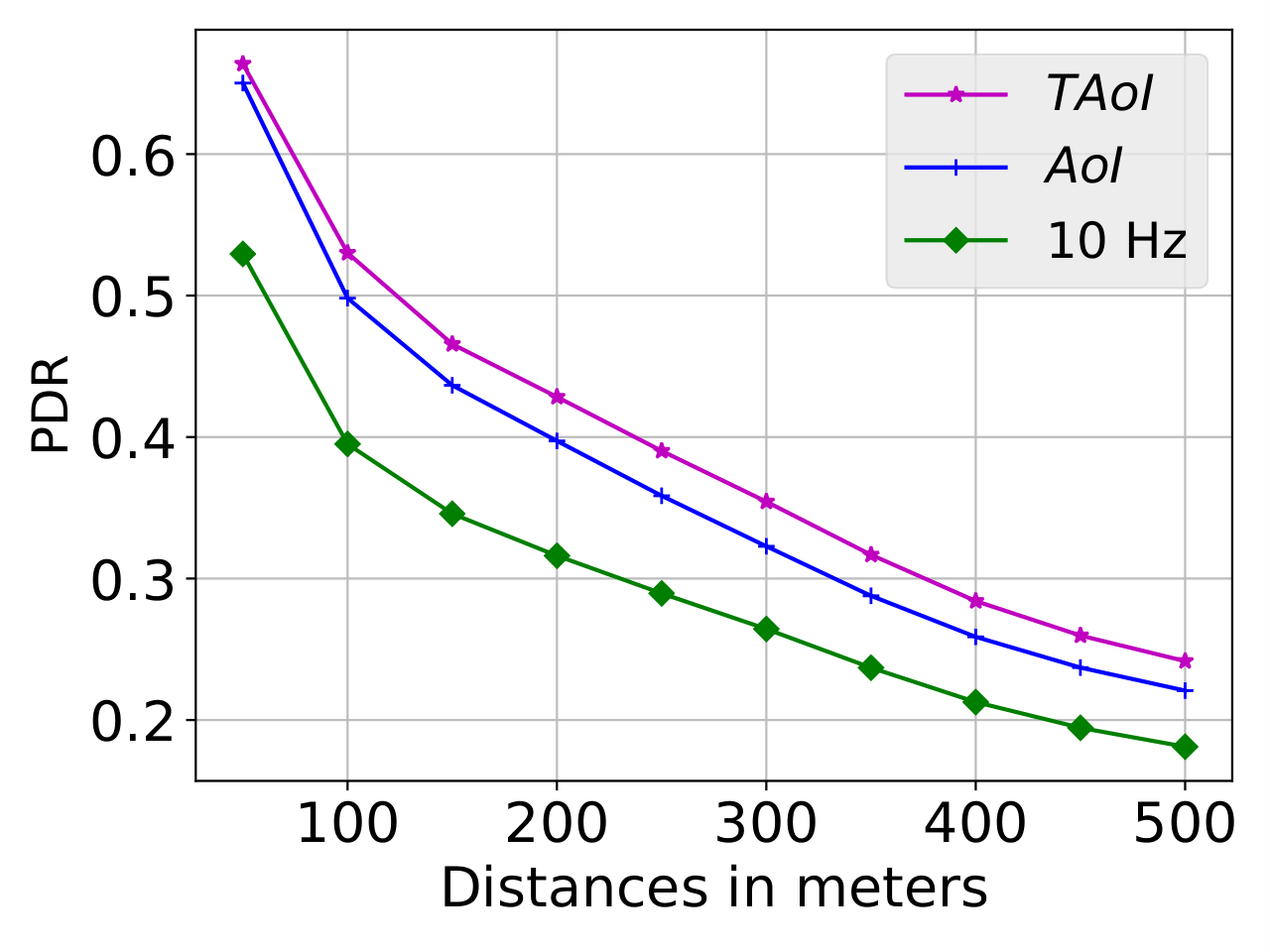,width=1.65 in}}
\caption{(a) Road Layout with a zoomed in picture showing cars moving in 3 lanes, (b) Collision Risk for different broadcast modes, (c) Density plot showing distribution of broadcast intervals, (d) Variation of PDR with distance}
\vspace{-0.15in}
\end{figure*}
\paragraph{Simulation Setting}
\label{para:simulation}
The 802.11p V2V network is simulated using network simulator-3 (ns-3) \cite{ns3_site}; and SUMO traffic simulation \cite{sumo_site} package is used to model the vehicle's mobility as it provides realistic mobility traces for V2V networks. A small part of the road simulated in SUMO with the vehicles in it is shown in Fig. \ref{fig:Road_SUMO} where the yellow triangles represent the vehicles. Unless otherwise stated, all the experiments are performed with 150 vehicles moving in a 3-lane rectangular road of 1000m $\times$ 100m. The total simulation time is 100 seconds. Table \ref{table - simulation-parameters} lists other important parameters.

\begin{table}[htb]
 \centering
 \small
 \vspace{-0.1in}
 \caption{Simulation Parameters}
    \begin{tabular}{|c|c|}
    \hline
    \textbf{Parameters}  & \textbf{Value} \\ \hline
    Number of vehicles ($\mathcal{N}$) & 100 - 250 \\ \hline
    Number of Lanes ($m$) & 3 \\ \hline
    Lane Width  & 4 m \\ \hline
    BSM size & 1000 bytes \\ \hline
    Transmission Power & 20 dBm \\ \hline
    Data Rate & 6 Mbps \\ \hline
    Loss Model & Log Distance Propagation Loss \cite{stoffers2012comparing}\\ \hline
    Path Loss exponent & $\gamma$ = 3  \\ \hline
    Fading & Nakagami-m \cite{stoffers2012comparing}  \\ \hline
    Channel Frequency & 5.9 GHz \\ \hline
    Channel Bandwidth & 10 MHz \\ \hline
    Max velocity ($s_{max}$) & 25 m/s \\ \hline
    Antennas/Spatial Streams & 1/1 \\ \hline
    $\tau_{p,th}$ & 0.5 m \\ \hline
    $t_{MI}$ & 1 s \\ \hline
    $\beta$ & 1.1 \\ \hline

 \end{tabular}
 \label{table - simulation-parameters}
\end{table}

\paragraph{Performance metrics.} For the comparative analysis of proposed $\AOIT$ rate control algorithm against the baseline $AoI$ rate control protocol and standard $10$ Hz, we consider the following two key performance metrics.

\begin{itemize}

     \item \textbf{Collision Risk} - It measures the overall on-road safety of V2V networks, and is defined as the number of instances between each pair of vehicles in which $\delta TTC$ exceeded $\delta TTC_{thres}$. Revisit Sec. \ref{subsubsection - Collision Risk} for details. 
    \item \textbf{Packet Delivery Ratio (PDR)} - It is the probability that all vehicles within the range of transmitting vehicle, successfully receives the transmitted BSM. It is defined as $PDR_v = \frac{PR_v}{PD_{v}}$ where $PD_v$ is number of BSMs sent by vehicle $v$ and $PR_v$ is the number of BSMs sent by $v$ received by neighboring vehicles $u \in \mathcal{N}_v$.

\end{itemize}

\begin{figure}
\vspace{-0.05in}
\subfigure[\label{fig:self_TE}]{
\epsfig{figure=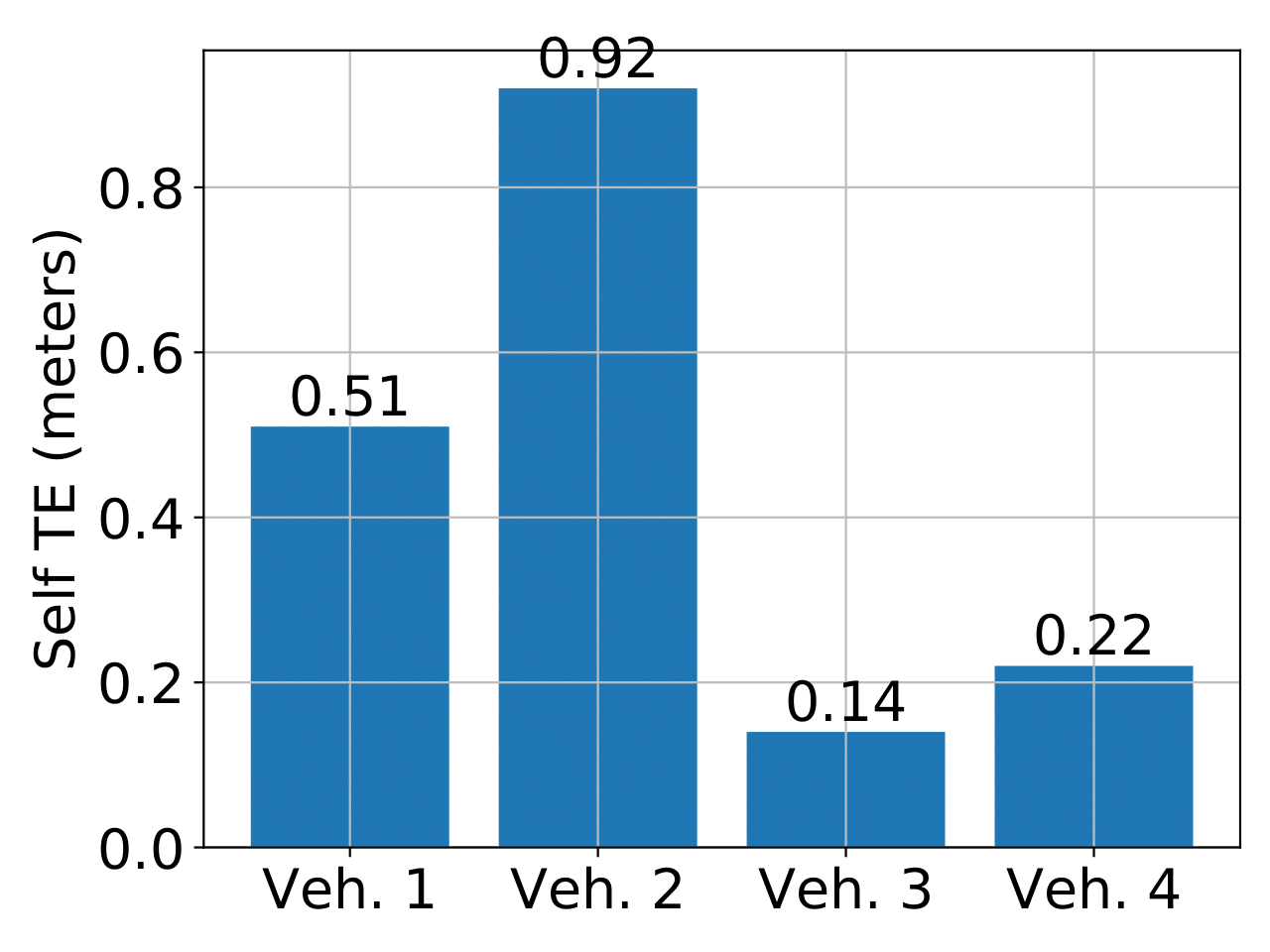,width=1.6 in,  keepaspectratio}}
\subfigure[\label{fig:self_TE_Interval}]{
\epsfig{figure=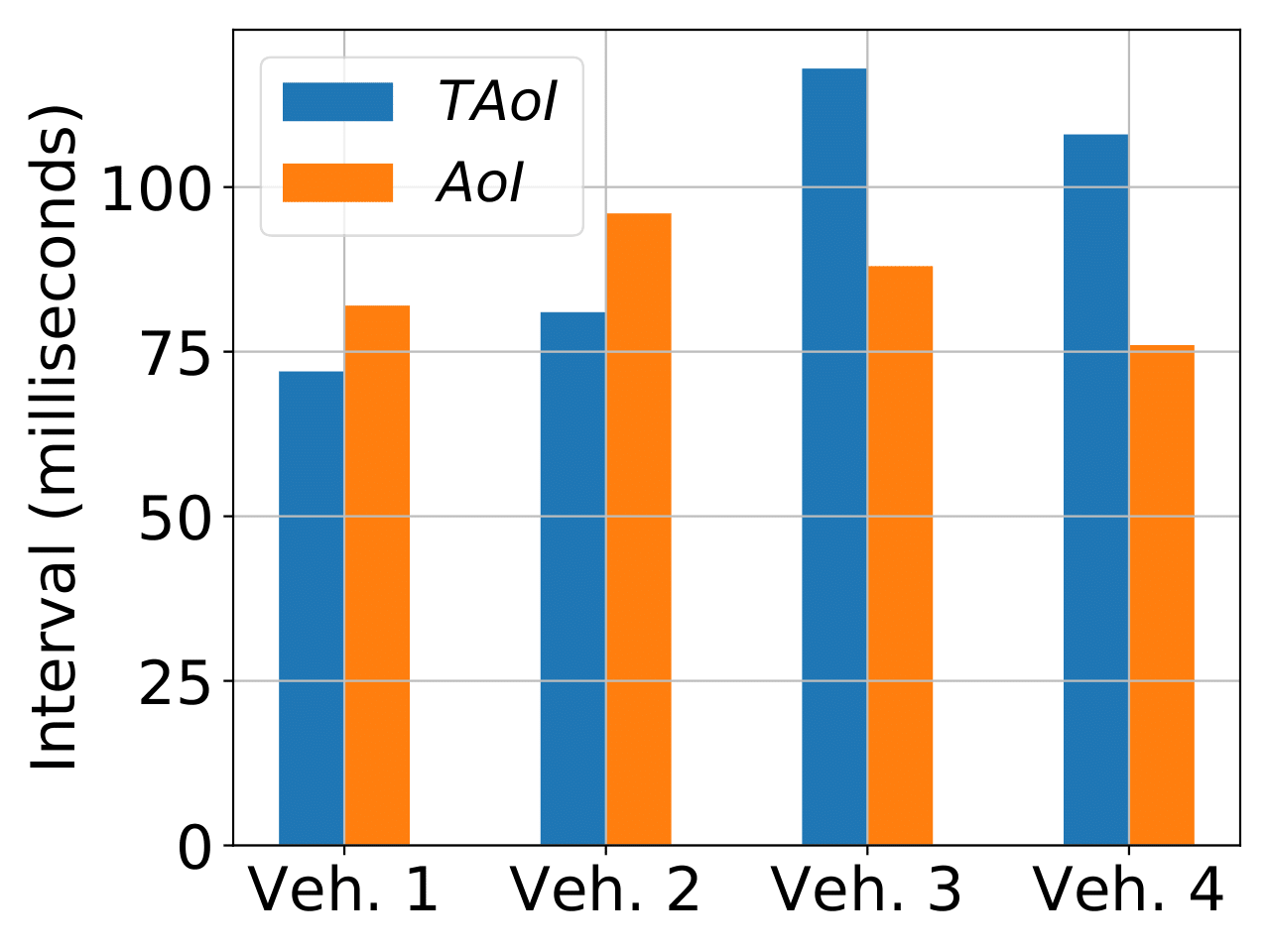,width=1.6 in,  keepaspectratio}}
\caption{ (a) Self TE as measured by 4 vehicles (b) Broadcast interval of the 4 vehicles}\vspace{-0.2in}
\end{figure}

\paragraph{Experimental Results}
\label{subsection - Experimental Results}

Before we present the results, we show the broadcast intervals selected by four different vehicles. As seen in Fig. \ref{fig:self_TE}, the first 2 vehicles are risky as their self-TE are (0.51m, 0.92m) greater than $\tau_{p,th}$. So $\AOIT$ prioritizes them by assigning them lower intervals compared to $AoI$ protocol as seen in \ref{fig:self_TE_Interval}. The opposite case can be seen for vehicles 3 and 4 which are non-risky, due to which the proposed $\AOIT$ protocol allocates it higher intervals compared to $AoI$ rate protocol.

The collision risk, distribution of the broadcast intervals and PDR for 150 vehicles are shown in Fig. \ref{fig:150U_Risk}, \ref{fig:150U_Intervals} and \ref{fig:150U_PDR}. From Fig. \ref{fig:150U_Risk}, it can be seen that both $AoI$ and $\AOIT$ improves safety compared to the default 10Hz broadcasting, which points to an strong relation between safety performance and improvement in $AoI$. Regarding the comparison between $AoI$ and $\AOIT$, the improvement in safety as as result of $\AOIT$ prioritizing the risky vehicles can be seen in Fig. \ref{fig:150U_Risk}. This shows that non-risky vehicles do not pose much of a threat in V2V networks. To show the different broadcast intervals for different vehicles, the density plot of the distribution of the broadcast intervals is then shown in Fig. \ref{fig:150U_Intervals} - it can be seen that both the protocols have a majority of their vehicles broadcasting around similar broadcast intervals. Due to the presence of a spread control factor\footnote{
As system $AoI$ was shown to be minimized only when each vehicle broadcasts at the same interval in \cite{kaul2011minimizing}, spread control is a necessary operation in system $AoI$ minimization presented there.} in \cite{kaul2011minimizing}, the intervals for $AoI$ protocol form a roughly symmetric distribution around the average interval. But with $\AOIT$ protocol, as each vehicle can independently select any interval without any spread control, the intervals are spread out with a few vehicles transmitting at very high intervals. Additionally the intervals are higher in our algorithm as vehicles increase their interval as soon as congestion is detected, which is not done in the case of \cite{kaul2011minimizing}. As a result of the higher intervals in $\AOIT$, channel congestion reduces and this results in a better PDR for $\AOIT$, as shown in Fig. \ref{fig:150U_PDR}. Hence the results show the benefit of a $\AOIT$ rate protocol as compared to plain $AoI$.

\begin{figure}
\vspace{-0.05in}
\subfigure[\label{fig:scalability_intervals}]{
\epsfig{figure=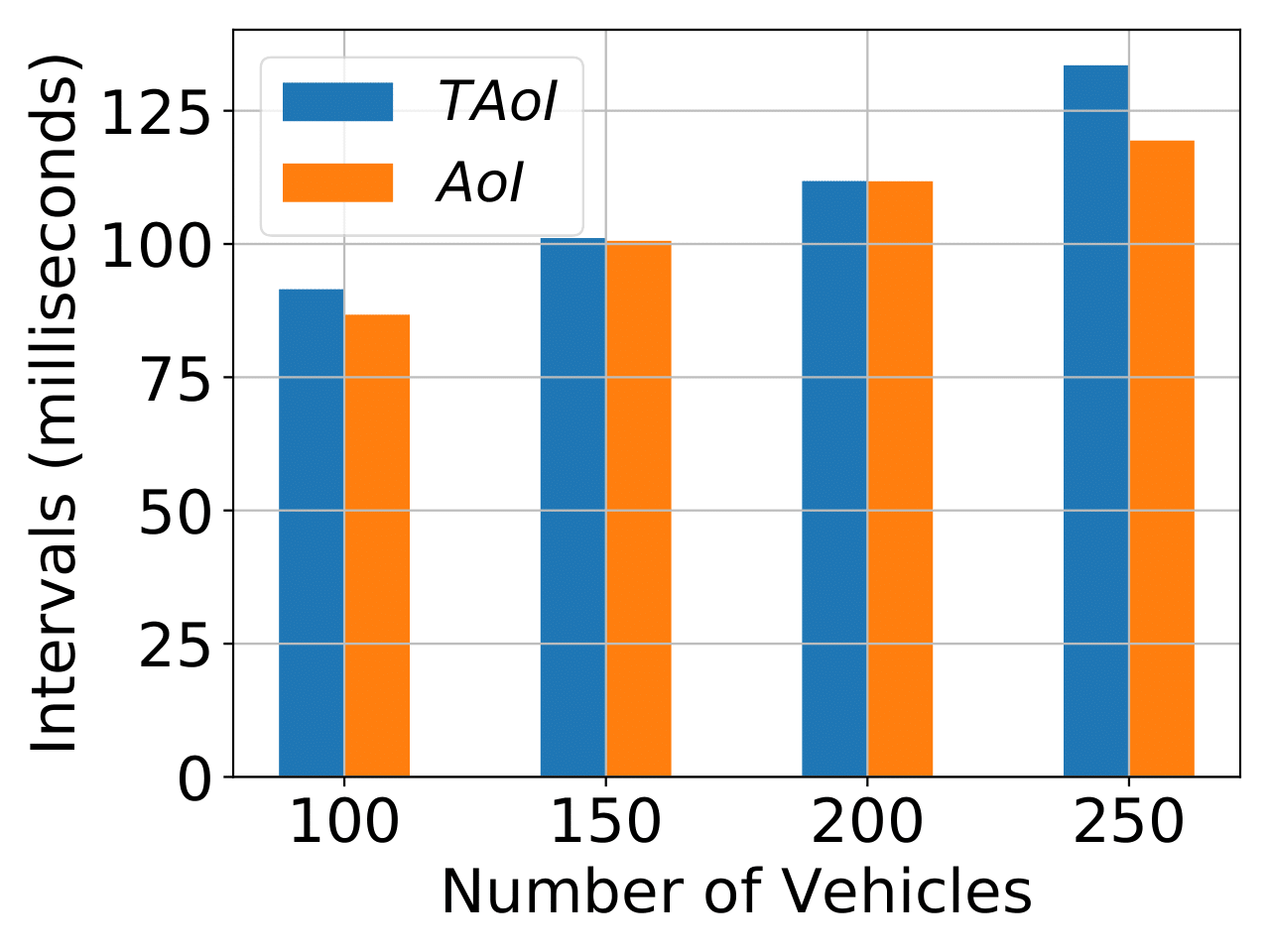,width=1.6 in,  keepaspectratio}}
\subfigure[\label{fig:scalability_risk}]{
\epsfig{figure=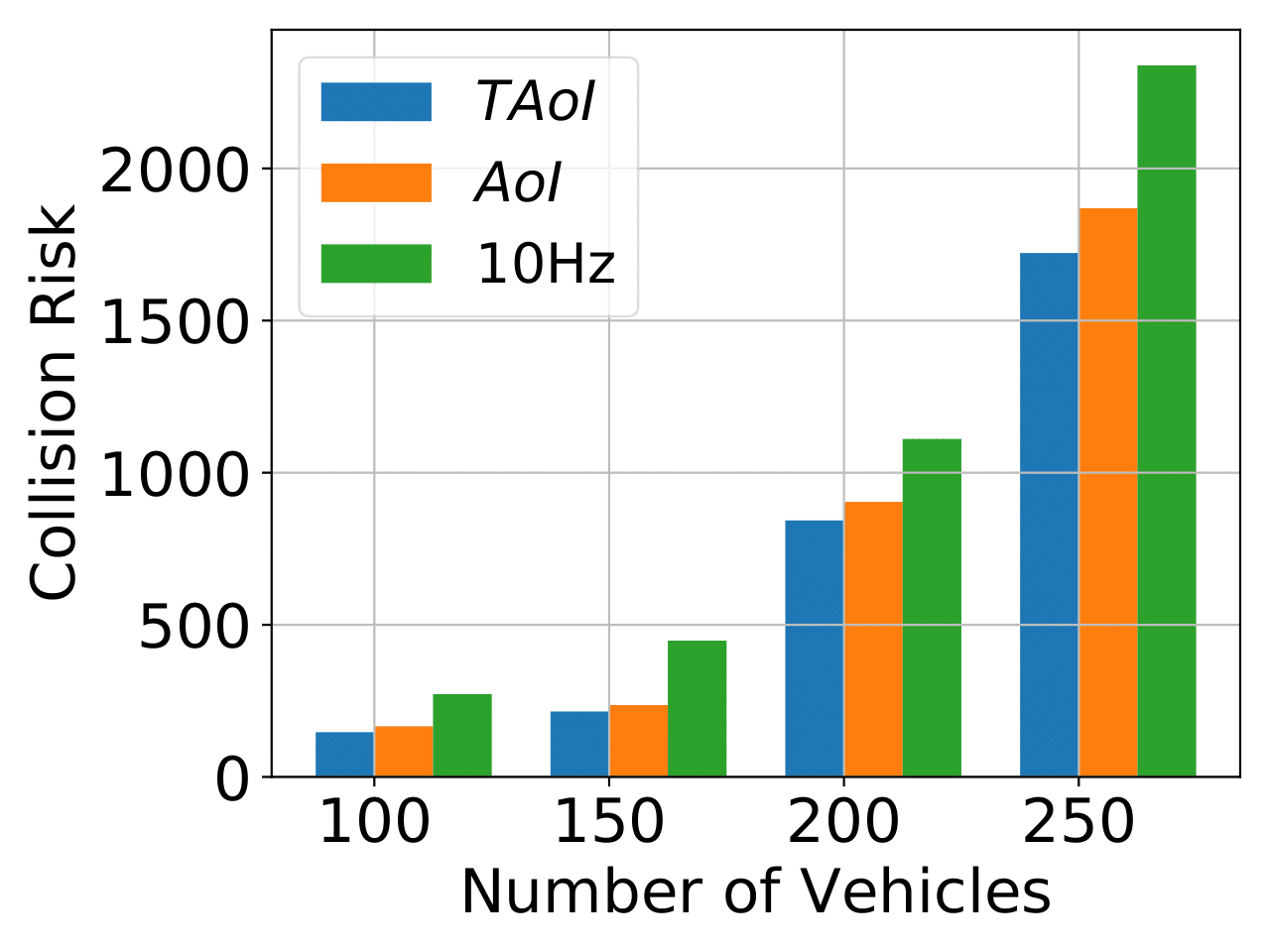,width=1.6 in,  keepaspectratio}}
\caption{ (a) Average broadcast intervals for different vehicle densities, (b) Collision Risk for different vehicle densities}
\vspace{-0.25in}
\end{figure}

\paragraph{Scalability Analysis} To see how the collision risk of the algorithm scales with the number of vehicles, results for 100, 150, 200 and 250 vehicles 
are shown. Their average broadcast intervals along with the collision risk are shown in Fig. \ref{fig:scalability_intervals} and Fig. \ref{fig:scalability_risk} respectively. For 100 vehicles, both the algorithms have an average interval of lower than 100ms, with average interval around 100ms for 150 vehicles and an average interval of more than 100ms for both 200 and 250 vehicles. This is consistent with the observation in \cite{choudhury2020experimental} where it has been shown that lower densities have lower average intervals that minimize the $AoI$ and vice versa. It is seen that the same relationship between the interval and density holds for $\AOIT$.

It is important to note from Fig. \ref{fig:scalability_intervals} and \ref{fig:scalability_risk} that $\AOIT$ consistently improves collision risk across different vehicle densities, while keeping the average broadcast intervals higher or equal to the intervals in $AoI$ rate protocol. The improvement in collision risk ranges around 8\%-12\% and 24\%-40\% compared to the $AoI$ rate protocol and default 10Hz broadcasting respectively. This result comprehensively proves that $\AOIT$ is a much better metric for controlling the broadcast interval - it utilizes the channel resources much efficiently as compared to $AoI$ rate protocol which is evident from the better safety performance while maintaining the broadcast intervals at same or higher levels compared to $AoI$ rate protocol.

\section{Conclusion}
\label{section - Conclusion}

In this paper, we showed that $AoI$ based rate control protocol does not always improve on-road safety in 802.11p V2V networks. To address this, we proposed a novel metric, termed, \textit{Trackability-aware} Age of Information, termed $\AOIT$, that jointly takes into account the $AoI$ and self risk assessment of vehicles, measured as self tracking error. Following this, we propose a decentralized $\AOIT$ based rate protocol for V2V networks that attempts to minimize locally measured $\AOIT$ at each vehicle, which in turn, improve the on-road safety of the V2V network. Our extensive experiments based on realistic SUMO traffic traces on top of ns-3 simulator demonstrate that $\AOIT$ rate control protocol greatly improves on-road safety of V2V networks compared to that of $AoI$ rate control and standard $10$ Hz rate, in all considered V2V scenarios.



\ifCLASSOPTIONcaptionsoff
  \newpage
\fi

\bibliographystyle{IEEEtran}
\bibliography{mybib}

\end{document}